\documentclass{jpp}
\pdfoutput = 1 
\usepackage {graphicx}
\usepackage{graphicx}
\usepackage{epstopdf, epsfig}
\usepackage{hyperref}
\usepackage{amsmath,amsfonts,amssymb}
\usepackage{subfigure}
\allowdisplaybreaks

\begin{document}
\title{Macroscopic Stability of a Rapidly Rotating Isothermal Magnetic Mirror}
\author{R.~Fitzpatrick\footnote{rfitzp@utexas.edu} and D.M.~Testa }
\affiliation{Institute for Fusion Studies,  Department of Physics, University of Texas at Austin,  Austin TX 78712, USA}
\maketitle

\begin{abstract}
The macroscopic ideal-MHD stability of an axisymmetric mirror device with sonic levels of plasma rotation is analyzed by approximating
the plasma equilibrium as a rotating isothermal theta pinch possessing an artificial gravity. 
Starting from a set of single-fluid, ideal-MHD, isothermal fluid equations, a linear, second-order, eigenmode equation for the radial component of the Lagrangian fluid displacement  is derived which governs the
stability of the equilibrium to small perturbations in the case of an arbitrary plasma angular velocity profile. The stability of
the $m=1$ and $m=2$ modes is investigated. For the case of rigid plasma rotation, the plasma is found to be stable to these two modes provided that it
is sufficiently short in the axial direction. For the case of sheared plasma rotation, the $m=2$ mode is found to be difficult to stabilize. This is the case for
both monotonically decreasing rotation profiles, and profiles that peak off the magnetic axis. 
\end{abstract}

\section{Introduction}
Mirror devices are among the simplest magnetic confinement configurations employed in nuclear fusion research. Mirrors only require axisymmetric magnetic field-coils, and avoid the geometric complexity associated with toroidal confinement systems [\cite{post,ryutov}]. In particular, the fact that mirrors are {\em linear}\/
devices implies that they do not possesses transverse particle drifts, thus negating the need for a rotational transform of the
confining magnetic field, which constitutes a very major advantage over toroidal confinement devices.
The principal obstacle to confinement in  axisymmetric mirrors is the ideal flute-interchange instability, whose existence was first predicted by Rosenbluth \& Longmire (1957) [\cite{rosenbluth}], and subsequently confirmed experimentally by Ioffe and co-workers  [\cite{ioffe}].
Recently, renewed interest in mirror confinement has accompanied the development of strong high-temperature-superconducting magnets, which has enabled the construction of compact mirror devices, such as the Wisconsin High-Temperature-Superconductor Axisymmetric Mirror (WHAM), that  possess  very large mirror ratios [\cite{wham}]. 

Let $r$, $\theta$, $z$ be a conventional cylindrical coordinate system. Consider an axisymmetric  mirror device that is co-axial with the $z$-axis, and
is bounded by perfectly conducting end-plates located at $z=0$ and $z=L$. The equilibrium magnetic field
is of the form ${\bf B} = B_r(r,z)\,{\bf e}_r+ B_z(r,z)\,{\bf e}_z$. The equilibrium ion velocity is written ${\bf V}= r\,{\mit\Omega}_\theta(\alpha)\,{\bf e}_\theta$,
where ${\bf B}\cdot\nabla \alpha=0$ [\cite{ferraro}]. Here, $\alpha(r,z)$ labels magnetic ``flux-surfaces'' (i.e., open magnetic flux-tubes), and ${\mit\Omega}_\theta$ is the plasma
angular velocity. This angular velocity   is approximately constant on a given flux-surface, provided 
the ion gyro-radius is much smaller than the machine dimensions, and ${\mit\Omega}_\theta$ is much
smaller in magnitude than the ion gyro-frequency [\cite{fitz1}].  The latter constraint would limit the Mach number to be much less than about 300
in an axisymmetric mirror with a plasma temperature of 1 keV, a magnetic field-strength of 1 T, and a radius of 1 m.  In WHAM, a specified ${\mit\Omega}_\theta(\alpha)$ profile can be driven
in the plasma 
by electrically biasing individual flux-surfaces by means of differential voltages applied to the end-plates, which are segmented, with the different segments being 
electrically insulated from one another  [\cite{wham}].
Previous studies have shown that moderate shear in the ${\mit\Omega}_\theta(\alpha)$ profile can reduce or suppress flute-interchange modes, but that
larger  values of ${\mit\Omega}_\theta(\alpha)$ gives rise to centrifugal forces that provide an additional interchange drive [\cite{aydemir,hojo}].

Unfortunately, a mirror confinement device is intrinsically ``two-dimensional"---meaning that the equilibrium is only characterized by a
single ignorable coordinate, $\theta$. This greatly complicates the calculation of the macroscopic stability of the confined plasma. In this paper, 
we  investigate  the ideal-magnetohydrodynamical (MHD) stability of a mirror device via a simplified one-dimensional model
in which the device is approximated as a  theta pinch. The unfavorable magnetic field-line curvature of a real mirror device is
simulated by means of an artificial gravity. 
Thus, the equilibrium magnetic field and ion velocity are assumed to have the simplified forms ${\bf B} =  B_z(r)\,{\bf e}_z$
and ${\bf V}= r\,{\mit\Omega}_\theta(r)\,{\bf e}_\theta$, respectively. The aim of our investigation is to determine what type of ${\mit\Omega}_\theta(r)$ profile is most conducive to ideal-MHD stability. 

Incidentally, a tokamak is another example of a two-dimensional plasma equilibrium that only possesses a single ignorable coordinate---namely,
the toroidal angle, $\phi$. Nevertheless, progress can be made in understanding the stability of tokamak plasmas by adopting a simplified one-dimensional model in which
the plasma is represented as a periodic cylinder [\cite{fitz1a}]. Indeed, the ``cylindrical tokamak" model is the foundation upon which all investigations of
the macroscopic stability of tokamak plasmas are built. In fact, the very vocabulary of tokamak stability theory is based on this simplified model. It
is hoped that the theta pinch model introduced in this paper can play the same role for magnetic mirror research that the cylindrical tokamak
model has played for tokamak research. 

\section{Analysis}
\subsection{Model fluid equations}
Our model single-fluid  ideal-MHD equations are [\cite{fitz}]:
\begin{align}
\nabla\cdot{\bf B} &= 0,\label{e1}\\[0.5ex]
\nabla\times{\bf B}&= \mu_0\,{\bf J},\\[0.5ex]
\frac{\partial{\bf B}}{\partial t} &= ({\bf B}\cdot\nabla)\,{\bf V} - ({\bf V}\cdot\nabla)\,{\bf B} - (\nabla\cdot {\bf V})\,{\bf B},\\[0.5ex]
0&=\frac{\partial N}{\partial t} + {\bf V}\cdot\nabla N +N\,\nabla\cdot{\bf V},\label{e3}\\[0.5ex]
{\bf J}\times {\bf B} &=m_i\,N\left[\frac{\partial {\bf V}}{\partial t} + ({\bf V}\cdot\nabla){\bf V}\right] +2\,T\,\nabla N -g\,T\,N\,r\,{\bf e}_r.\label{e2.7}
\end{align}
Here, ${\bf B}$ is the magnetic field-strength, ${\bf J}$ the current density, ${\bf V}$ the MHD (i.e., guiding center) fluid
velocity, $N$ the particle number density, $m_i$ the ion mass,  and $T$ the common temperature of the electrons and ions. For the sake
of simplicity, $T$ is assumed to be spatially and temporally constant. 
The final term on the right-hand side of Eq.~(\ref{e2.7}) is an artificial gravity that represents the unfavorable average magnetic
curvature of magnetic field-lines in a conventional mirror device. Here, $\rho_c=(g\,r)^{-1}$ is the effective radius of curvature of the field-lines.
Note that the radius of curvature tends to infinity as $r\rightarrow 0$ because the magnetic field-lines on the axis of a mirror machine are straight. 

Our isothermal (i.e., ${\mit\Gamma}=1$, where ${\mit\Gamma}$ is the polytropic index [\cite{chand}]) approximation is inspired by the fact that, in the type of hot, essentially collisionless plasmas found in mirror devices, electron heat is convected along equilibrium magnetic field-lines
at the electron thermal speed, which implies that any perturbations in the electron temperature along  field-lines are smoothed out by the large parallel electron heat transport
far faster than they can be generated by an MHD instability. For the case of ions, the situation is less clear, because ion heat is convected along magnetic field-lines at the ion thermal
speed, which implies that any perturbations in the ion temperature  along  field-lines are smoothed out about as fast as they are 
 generated by an MHD instability. As is usually the case when dealing with the macroscopic stability of an essentially collisionless, highly magnetized plasma, we expect a
fluid approach to do a fairly good job of modeling those aspect of instabilities associated with displacements perpendicular to magnetic field-lines, but a fairly
poor job of modeling those aspects associated with parallel displacements [\cite{fitz1a}]. Ultimately, the deficiencies of the fluid approach can only be corrected by
adopting a kinetic model of the parallel dynamics [\cite{rosen}]. However, in practice, this so-called ``kinetic-MHD''  approach is prohibitively difficult
to implement. 

\subsection{Normalization}
Let $a$ be a typical equilibrium scale-length, $B_0$  a typical magnetic field-strength, and $N_0$ a typical particle
number density. The mean ion thermal speed is $v_i=(T/m_i)^{1/2}$. 
Let $ t=(a/v_i)\,\hat{t}$, $\hat{r}=r/a$, $\hat{z}=z/a$, $\hat{L}=L/a$, $\nabla = a^{-1}\,\hat{\nabla}$, $g=\hat{g}/a^{\,2}$, $N= N_0\,\hat{N}$, ${\bf V} = v_i\,\hat{\bf V}$, ${\bf B}= B_0\,\hat{\bf B}$,  and ${\bf J} =(B_0/\mu_0\,a)\,\hat{\bf J}$. 
 It follows that 
\begin{align}
\hat{\nabla}\cdot\hat{\bf B} &= 0,\\[0.5ex]
\hat{\nabla}\times\hat{\bf B}&= \hat{\bf J},\\[0.5ex]
\frac{\partial\hat{\bf B}}{\partial \hat{t}} &=(\hat {\bf B}\cdot\hat{\nabla})\,\hat{\bf V} - (\hat{\bf V}\cdot\hat{\nabla})\,\hat{\bf B}-(\hat{\nabla}\cdot\hat{\bf V})\,\hat{\bf B},\\[0.5ex]
0&=\frac{\partial \hat{N}}{\partial \hat{t}} + \hat{\bf V}\cdot\hat{\nabla} \hat{N} +\hat{N}\,\hat{\nabla}\cdot\hat{\bf V},\label{epre}\\[0.5ex]
\hat{\bf J}\times\hat{\bf B} &=\beta_0\left(\hat{N}\left[\frac{\partial\hat{\bf V}}{\partial\hat{t}}+ (\hat{\bf V}\cdot\hat{\nabla})\,\hat{\bf V}\right]
+2\,\hat{\nabla}\hat{N} -\hat{N}\,\hat{g}\,\hat{r}\,{\bf e}_r\right),
\end{align}
where 
\begin{equation}
\beta_0 = \frac{\mu_0\,N_0\,T}{B_0^{\,2}}
\end{equation}
is a typical ratio of the plasma thermal energy density to the magnetic energy density. 

\subsection{Plasma equilibrium}
Suppose that the plasma  equilibrium is characterized by 
$\hat{\bf B} = \hat{B}_z(\hat{r})\,{\bf e}_z$, 
$\hat{\bf J} = \hat{J}_\theta(\hat{r})\,{\bf e}_\theta$, 
$\hat{\bf V} = \hat{r}\,\hat{\mit\Omega}_\theta(\hat{r})\,{\bf e}_\theta$, 
and $\hat{N} = \hat{N}(r)$. 
It follows that
\begin{align}\label{equil}
\hat{J}_\theta &=-\hat{B}_z',\\[0.5ex]
\beta_0\,\hat{N}\,\hat{r}\,(\hat{g}+\hat{\mit\Omega}_\theta^{\,2})&= \frac{d}{d\hat{r}}\!\left(\frac{\hat{B}_z^{\,2}}{2} + 2\,\beta_0\,\hat{N}\right),\label{equil1}
\end{align}
where $'\equiv d/d\hat{r}$.

It can be seen that, in the absence of an equilibrium plasma current (i.e., $\hat{J}_\theta=0$), both the artificial
gravity (proportional to $\hat{g}$) and the centrifugal force (proportional to $\hat{\mit\Omega}_\theta^{\,2}$) would support a
positive equilibrium density gradient (i.e., $\hat{N}'>0$). The fact that the density gradient is actually  negative in a conventional
equilibrium possessing a plasma current sets up the so-called ``adverse'' pressure gradient that is the drive for the flute-interchange mode. 

Suppose that, in the mirror device that our theta pinch is modeling, the magnetic field-line that passes through the point $r=r_0$, $z=L/2$ has the
equation 
\begin{equation}
r = \frac{4\,r_0}{L^{\,2}}\,z\,(L-z).
\end{equation}
This equation, which is only approximate,  is consistent with a mirror device, such as WHAM,  that possesses a very large mirror ratio. If we identify the mean radius of curvature of the
field-line, $\rho_c$, with the radius of curvature at the midpoint then
\begin{equation}
\rho_c = \left(\frac{1}{|d^{\,2}r/dz^{\,2}|}\right)_{z=L/2}=\frac{L^{\,2}}{8\,r_0}=\frac{1}{g\,r_0}.
\end{equation}
Hence, we deduce that
\begin{equation}
\hat{g} = \frac{8}{\hat{L}^{\,2}}.
\end{equation}

\subsection{Eigenmode equation}
Consider a small perturbation to the plasma equilibrium. Let $\delta\hat{\bf B}$ be the perturbed magnetic field, et cetera. The linearized equations governing the
perturbation are:
\begin{align}\label{e4.1}
\hat{\nabla}\cdot\delta\hat{\bf B} &= 0,\\[0.5ex]
\hat{\nabla}\times\delta\hat{\bf B}&= \delta\hat{\bf J},\label{e4.3}\\[0.5ex]
\frac{\partial\delta\hat{\bf B}}{\partial \hat{t}}&= (\hat{\bf B}\cdot\hat{\nabla})\,\delta\hat{\bf V} 
+(\delta\hat{\bf B}\cdot\hat{\nabla})\hat{\bf V}-(\delta\hat{\bf V}\cdot\hat{\nabla})\hat{\bf B} -(\hat{\bf V}\cdot\hat{\nabla})\,\delta\hat{\bf B}\nonumber\\[0.5ex]
&\phantom{=}-(\hat{\nabla}\cdot\delta\hat{\bf V})\,\hat{\bf B},\\[0.5ex]
0&=\frac{\partial \delta\hat{N}}{\partial \hat{t}} + \delta\hat{\bf V}\cdot\hat{\nabla} \hat{N}+ \hat{\bf V}\cdot\hat{\nabla} \delta\hat{N} +(\hat{\nabla}\cdot\delta\hat{\bf V})\,\hat{N},\\[0.5ex]
\delta\hat{\bf J}\times\hat{\bf B}+\hat{\bf J}\times\delta\hat{\bf B}&=\beta_0\,\delta\hat{N}\,(\hat{\bf V}\cdot\hat{\nabla})\,\hat{\bf V}+2\,\beta_0\,\hat{\nabla}\delta\hat{N}-\beta_0\,\hat{g}\,\hat{r}\,\delta\hat{N}\,{\bf e}_r\nonumber\\[0.5ex]
&\phantom{=}+ \beta_0\,\hat{N}\left[\frac{\partial\delta\hat{\bf V}}{\partial\hat{t}}+ (\delta\hat{\bf V}\cdot\hat{\nabla})\hat{\bf V}+ (\hat{\bf V}\cdot\hat{\nabla})\,\delta\hat{\bf V}\right]\label{e4.3d}.
\end{align}

Suppose that all perturbed quantities are of the form
\begin{equation}\label{e4.7}
\delta a(\hat{r},\theta,\hat{z},\hat{t}) =
 \delta a(\hat{r})\,\exp\left[\,{\rm i}\left(m\,\theta + \hat{k}\,\hat{z} - \hat{\omega}\,\hat{t}\right)\right],
\end{equation}
where $m$ is an integer. 
Let
\begin{align}
\sigma(\hat{r}) &= \hat{\omega}- m\,\hat{\mit\Omega}_\theta(\hat{r}),\label{esig}\\[0.5ex]
\beta(\hat{r})& = \frac{\beta_0\,\hat{N}}{\hat{B}_z^{\,2}},\label{betadef}\\[0.5ex]
F_a(\hat{r}) &= \hat{k}^2-\beta\,\sigma^2,\label{fadef}\\[0.5ex]
F_s(\hat{r}) &= 2\,\hat{k}^{\,2}-\sigma^2,\label{fsdef}\\[0.5ex]
F_f(\hat{r}) &= 2\,\hat{k}^{\,2} - (1+2\,\beta)\,\sigma^2.\label{ffdef}
\end{align}
Note that $\sigma$ is the complex frequency of the perturbation in a frame of reference that locally co-rotates with the plasma, $F_a=0$ is the dispersion relation of a shear-Alfv\'{e}n wave that propagates perpendicular to magnetic field-lines,
$F_s=0$ the dispersion relation of a sound wave that propagates along magnetic field-lines, and $F_f=0$ the
dispersion relation of a compressible-Alfv\'{e}n wave that propagates perpendicular to magnetic field-lines [\cite{fitz}].

After a great deal of analysis (see Sect.~\ref{secta}), Eqs.~(\ref{e4.1})--(\ref{e4.7}) yield the second-order, linear, eigenmode equation
\begin{equation}\label{eig}
\frac{d}{d\hat{r}}\!
\left(\frac{N}{D}\,\frac{d\zeta}{d\hat{r}}\right) = \left[A+\frac{B}{D}+\left(\frac{C}{D}\right)'\right]\zeta,
\end{equation}
where
\begin{align}
\zeta(\hat{r}) &= \frac{\hat{r}\,\delta\hat{V}_r}{\sigma},\\[0.5ex]
N(\hat{r}) &= \hat{B}_z^{\,2}\,\hat{r}\,F_a\,F_f,\\[0.5ex]
A(\hat{r}) &= \frac{\hat{B}_z^{\,2}\,F_a}{\hat{r}}+2\,\beta_0\,\hat{N}\,\hat{\Omega}_\theta\,\hat{\Omega}_\theta'\,\\[0.5ex]
B(\hat{r})&= \beta_0\,\hat{N}'\,(\hat{g}+\hat{\mit\Omega}_\theta^{\,2})\,[(m^2+\hat{k}^2\,\hat{r}^{\,2})\,F_f+\beta\,\sigma^4\,\hat{r}^{\,2}]
\nonumber\\[0.5ex]
&\phantom{=}-\beta_0\,\hat{N}\,\hat{r}\,(\hat{g}+\hat{\mit\Omega}_\theta^{\,2})^2\,(m^2+\hat{k}^2\,\hat{r}^{\,2})\,F_a+ 4\,\beta_0\,\hat{N}\,(\hat{g}+\hat{\Omega}_\theta^{\,2})\,\beta\,m\,\sigma^{3}\,\hat{r}\,\hat{\Omega}_\theta\nonumber\\[0.5ex]
&
\phantom{=}-4\,\beta_0\,\hat{N}\,\beta\,\sigma^2\,\hat{r}\,\hat{\Omega}_\theta^{\,2}\,F_s,\\[0.5ex]
C(\hat{r})&= \beta_0\,\hat{N}\,[\sigma^2\,\hat{r}^{\,2}\,(\hat{g}+\hat{\mit\Omega}_\theta^{\,2})\,F_a+2\,m\,\sigma\,{\mit\Omega}_\theta\,F_f],\\[0.5ex]
D(\hat{r}) &= m^2\,F_f+\hat{r}^{\,2}\,F_a\,F_s.
\end{align}
Note that $\xi(\hat{r})= \delta \hat{V}_r/\sigma$ is the radial component of the Lagrangian displacement of a fluid element, seen in a frame of
reference that locally co-rotates with the plasma. Moreover, $\zeta(\hat{r})=\hat{r}\,\xi$. 

Equation~(\ref{eig}) is different to the eigenmode equation derived in \cite{wesson}, because  the latter equation was derived under the
assumption that the plasma is incompressible, $\hat{\mit\Omega}_\theta$ is a spatial constant, and $\hat{g}=0$. Equation~(\ref{eig}) is  fully consistent with Eq.~(11.113) in \cite{freidberg}.
However, the latter equation was derived under the assumption that $\hat{\mit\Omega}_\theta = 0$ and $\hat{g}=0$. 
Although it is far from clear at first sight, our eigenmode equation is, in fact, equivalent to Eq.~(3) in \cite{bond} and  Eq.~(12) in \cite{goeda} (in the limit ${\mit\Gamma}=1$, $\hat{g}=0$,
 $\hat{B}_\theta=0$, and $\hat{V}_z=0$). The latter two equations were derived starting from the Frieman-Rotenberg equation [\cite{frie,goed}], rather than the more
 direct approach adopted in this paper. 
 For the sake of clarity, we have given the
key steps in the derivation of our  eigenmode equation in Sect.~\ref{secta}. 

\subsection{Boundary conditions}
If we restrict our attention to modes with non-zero poloidal mode numbers then, close to the magnetic axis, the well-behaved solution of the eigenmode equation, (\ref{eig}), is
such that
\begin{equation}
\zeta\propto \hat{r}^{\,|m|}.
\end{equation}

When written in terms of $\zeta(\hat{r})$, the components of the perturbed magnetic field take the forms (see Sect.~\ref{secta}):
\begin{align}
\frac{\delta\hat{B}_r}{\hat{B}_z}&= - \frac{\hat{k}}{\hat{r}}\,\zeta,\\[0.5ex]
\frac{{\rm i}\,\delta\hat{B}_\theta}{\hat{B}_z} &= \frac{\hat{k}\,m\,F_f}{D}\,\zeta' - \hat{k}\,\beta\,\hat{r}\left[\frac{m\,\sigma^2\,(\hat{g}+\hat{\Omega}_\theta^{\,2})-2\,\sigma\,\hat{\mit\Omega}_\theta\,F_s}{D}\right]\zeta,\\[0.5ex]
\frac{{\rm i}\,\delta\hat{B}_z}{\hat{B}_z} &= \frac{\hat{r}\,F_a\,F_s}{D}\,\zeta'+
\left(\frac{\hat{B}_z'}{\hat{r}\,\hat{B}_z} +m\,\beta\left[\frac{m\,\sigma^2\,(\hat{g}+\hat{\Omega}_\theta^{\,2})-2\,\sigma\,\hat{\mit\Omega}_\theta\,F_s}{D}\right]\right)\zeta. 
\end{align}

Suppose that the plasma is surrounded by a rigid, concentric, perfectly conducting wall located at $r=b$. Let $\hat{b}= b/a$. 
The physical  boundary conditions, $\delta \hat{V}_r = \delta\hat{B}_r=0$,  that must be imposed at the wall, imply that
\begin{equation}
\zeta(\hat{b})= 0.
\end{equation}

Note that the eigenmode equation, (\ref{eig}), only depends on $\hat{k}^{\,2}$. Thus, modes with the same poloidal mode number,
$m$, but equal and opposite values of $\hat{k}$, have the same values of $\hat{\omega}$. It follows that we can set up
standing wave patterns such that $\delta\hat{B}_r$ and $\delta\hat{B}_\theta$ vary in $\hat{z}$ as $\cos(|\hat{k}|\,\hat{z})$,
whereas $\delta\hat{B}_z$ varies as $\sin(|\hat{k}|\,\hat{z})$. The physical boundary conditions that must be imposed  at the perfectly conducting end-plates
are $\delta\hat{B}_z(z=0)=\delta\hat{B}_z(z=L)=0$. We can satisfy these boundary conditions by choosing 
\begin{equation}
\hat{k} = \frac{l\,\pi}{\hat{L}},
\end{equation}
where  $l$ is a positive integer. 

\subsection{Discussion}
Strictly speaking, a ``flute'' mode is characterized by $\hat{k}=0$. However, such modes are prohibited by the
boundary conditions imposed at the two end-plates. However, modes with finite values of $\hat{k}$  are closely related to true flute modes, in that they
are driven unstable by the same interchange mechanism.

Let
us write $\hat{\omega} = \hat{\omega}_r + {\rm i}\,\hat{\gamma}_r$, where $\hat{\omega}_r$ is the normalized real frequency of the
perturbation, and $\hat{\gamma}_r$ is the normalized real growth-rate. 
Suppose that $\zeta(\hat{r})$ is an eigenfunction of Eq.~(\ref{eig}) that satisfies the boundary conditions and  possess the eigenvalue $\sigma$. 
It is easily seen that $\zeta^\ast(r)$ is also an eigenfunction of Eq.~(\ref{eig}) that  that satisfies the boundary conditions and  possess the eigenvalue $\sigma^\ast$. Thus, if a mode exists with real frequency $\hat{\omega}_r$ and real growth-rate  $\hat{\gamma}_r$ then a
corresponding mode exists with real frequency $\hat{\omega}_r$ and real growth-rate  $-\hat{\gamma}_r$. 
Furthermore, if $m\rightarrow -m$ and $\hat{\omega}\rightarrow-\hat{\omega}$ then $\sigma\rightarrow-\sigma$ and $\zeta\rightarrow\zeta$. 
Thus, if an unstable mode exists with poloidal mode number $m$, real frequency $\hat{\omega}_r$, and real growth-rate $\hat{\gamma}_r$,
then a corresponding mode exists with poloidal mode number $-m$, real frequency $-\hat{\omega}_r$, and real growth-rate $\hat{\gamma}_r$. 
It follows that modes with negative poloidal mode numbers are equivalent to modes with the corresponding positive mode numbers,
except that they rotate in the opposite direction. Consequently, we can find modes in which $\delta\hat{B}_r$ varies as $\cos(|m|\,\theta-\hat{\omega}_r\,\hat{t})\,\cos(|\hat{k}|\,\hat{z})$, $\delta\hat{B}_\theta$ varies as $\sin(|m|\,\theta-\hat{\omega}_r\,\hat{t})\,\cos(|\hat{k}|\,\hat{z})$
 and $\delta\hat{B}_z$ varies as 
vary as $\cos(|m|\,\theta-\hat{\omega}_r\,\hat{t})\,\sin(|\hat{k}|\,\hat{z})$, where $\hat{\omega}_r$
is the real frequency associated with the positive poloidal mode number. 

\subsection{Classification of modes}
A macroscopic instability is characterized by three mode numbers: $n$, $m$, and $l$. 
Here, modes with increasing $n$ numbers have an increasing number of oscillations in the radial direction, with an $n=0$ mode having the least
number. Moreover,  $m$ is the number of nodes in the poloidal direction. Finally, $l-1$ is the number of nodes in the axial direction, not
counting the nodes at the two end-plates. For a
given value of $m$, the most unstable mode is always found to correspond to $n=0$ and $l=1$. Let us investigate the stability of the two
most dangerous macroscopic modes: namely, the $n=0$, $m=1$, $l=1$ mode and the $n=0$, $m=2$, $l=1$ mode. 

\section{Rigid rotation}\label{rigid}
\subsection{Equilibrium}
Suppose  that the plasma is rotating as a rigid body, so that $\hat{\mit\Omega}_\theta$ is independent of $\hat{r}$.
Let us adopt  the rigid rotator equilibrium of \cite{wesson}:
\begin{align}
\hat{N}(\hat{r}) &= \frac{1}{\cosh^2(\hat{r}^{\,2})},\\[0.5ex]
\lambda_0 &= \frac{1}{8}\,(\hat{g}+\hat{\mit\Omega}_\theta^{\,2}),\\[0.5ex]
\alpha_0&=\frac{\lambda_0}{1+\lambda_0},\label{e7.4}\\[0.5ex]
\beta_0&= \frac{1}{4}\,(1-\alpha_0)^2,\\[0.5ex]
\hat{B}_z(\hat{r})& =\alpha_0 + (1-\alpha_0)\,\tanh(\hat{r}^{\,2}),\\[0.5ex]
\hat{J}_\theta(\hat{r})&= -\frac{2\,\hat{r}\,(1-\alpha_0)}{\cosh^2(\hat{r}^{\,2})}.
\end{align}
These equations are exact solutions of Eqs.~(\ref{equil}) and (\ref{equil1}). Note that 
$a$ is a
measure of the width of the current channel, 
$N_0$ the central electron number density, $B_0$  the vacuum magnetic field-strength, and the beta value at the center of the plasma is $\beta_0/\alpha_0^{\,2}$. 
Incidentally, other rigid rotator equilibria are possible, but
they all possess lower values of $\beta_0$ than the one adopted in this paper [\cite{morse}]. 

\begin{figure}
\centering
\includegraphics[width=0.9\textwidth]{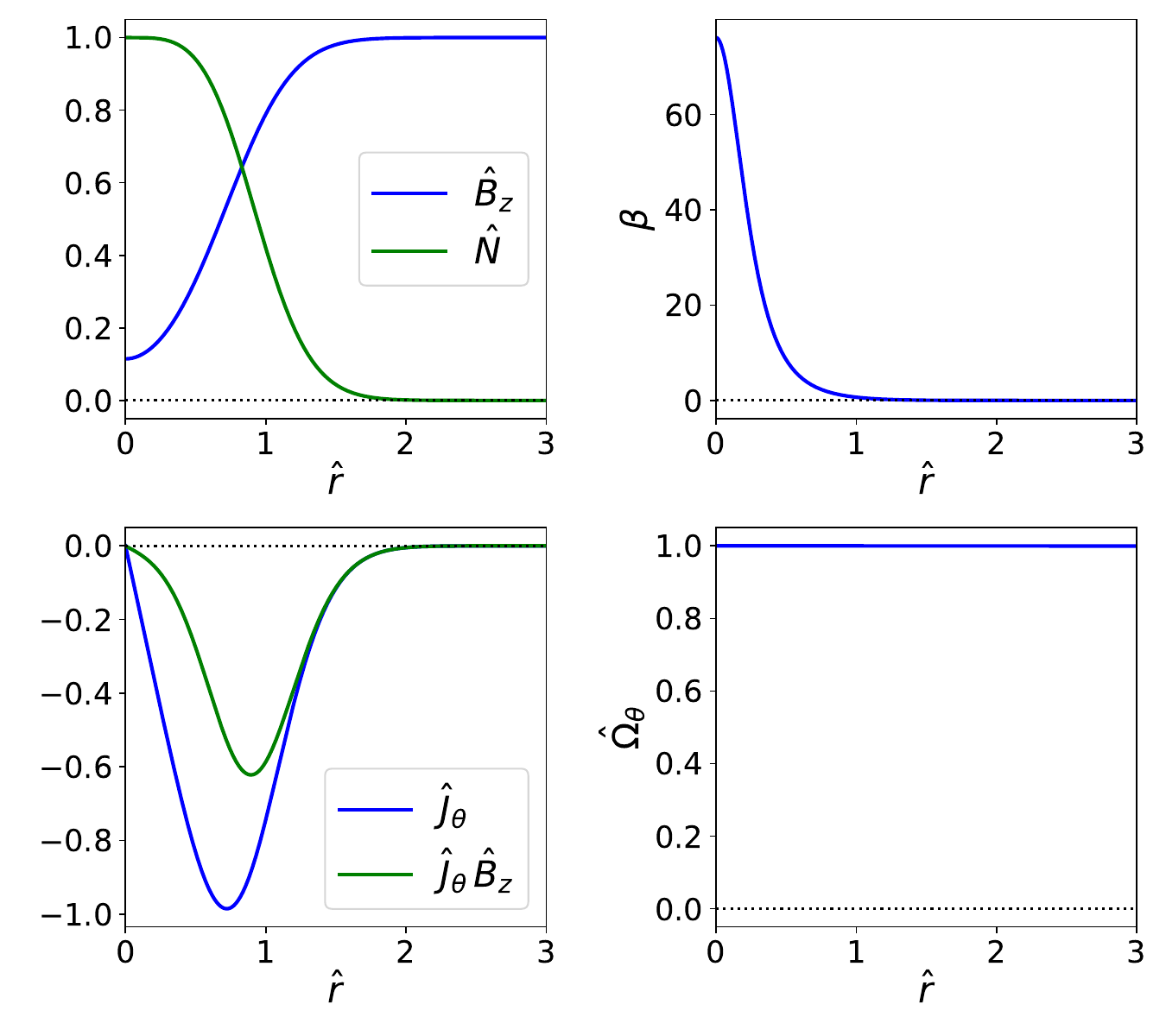}
\caption{A rigidly rotating plasma equilibrium calculated for $\hat{L}=15.0$ and $\hat{\mit\Omega}_\theta=1.0$.}\label{fig1}
\end{figure}

Figure~\ref{fig1} shows an example rigidly rotating plasma
equilibrium calculated for $\hat{L} = 15.0$ and $\hat{\mit\Omega}_\theta=1.0$. The parameters that characterize this equilibrium are
$\lambda_0= 0.129$, $\alpha_0=0.115$, and $\beta_0= 0.196$. The average $\beta$ value, which is defined as $\langle \beta\rangle \equiv \beta_0\,\langle \hat{N}/\hat{B}_z^{\,2}\rangle$, where $\langle\cdots\rangle$ denotes a volume average, is $\langle \beta\rangle= 0.190$. 

\subsection{Eigenmode equation}
 We can determine
$\hat{\omega}_r$ and $\hat{\gamma}_r$ by solving the following pair of complex first-order ordinary differential equations [see Eq.~(\ref{eig})]:
\begin{align}\label{e9.1}
\frac{d\zeta}{d\hat{r}} &=\frac{D}{N}\,\chi,\\[0.5ex]
\frac{d\chi}{d\hat{r}} &= \left[A+ \frac{B}{D}+\left(\frac{C}{D}\right)'\right]\zeta,\label{e9.2}
\end{align}
The boundary conditions at small $\hat{r}$ are  $\zeta=\hat{r}^{\,|m|}$ and $\chi = (\hat{B}_z^{\,2}\,F_a/|m|)_{\hat{r}=0}\,\hat{r}^{\,|m|}$.
The boundary condition at $\hat{r}=\hat{b}$ is $\zeta(\hat{b})=0$. 
Well-behaved  solutions of Eqs.~(\ref{e9.1}) and (\ref{e9.2}) must be launched from the magnetic axis, and integrated
 to the plasma boundary. This process must be repeated, for different values of $\hat{\omega}_r$ and $\hat{\gamma}_r$, until the real and imaginary parts of $\zeta(\hat{b})$ are simultaneously set to zero. 
 
\subsection{The $n=0$, $m=1$, $l=1$ mode}
\begin{figure}
\centering
\includegraphics[width=1.0\textwidth]{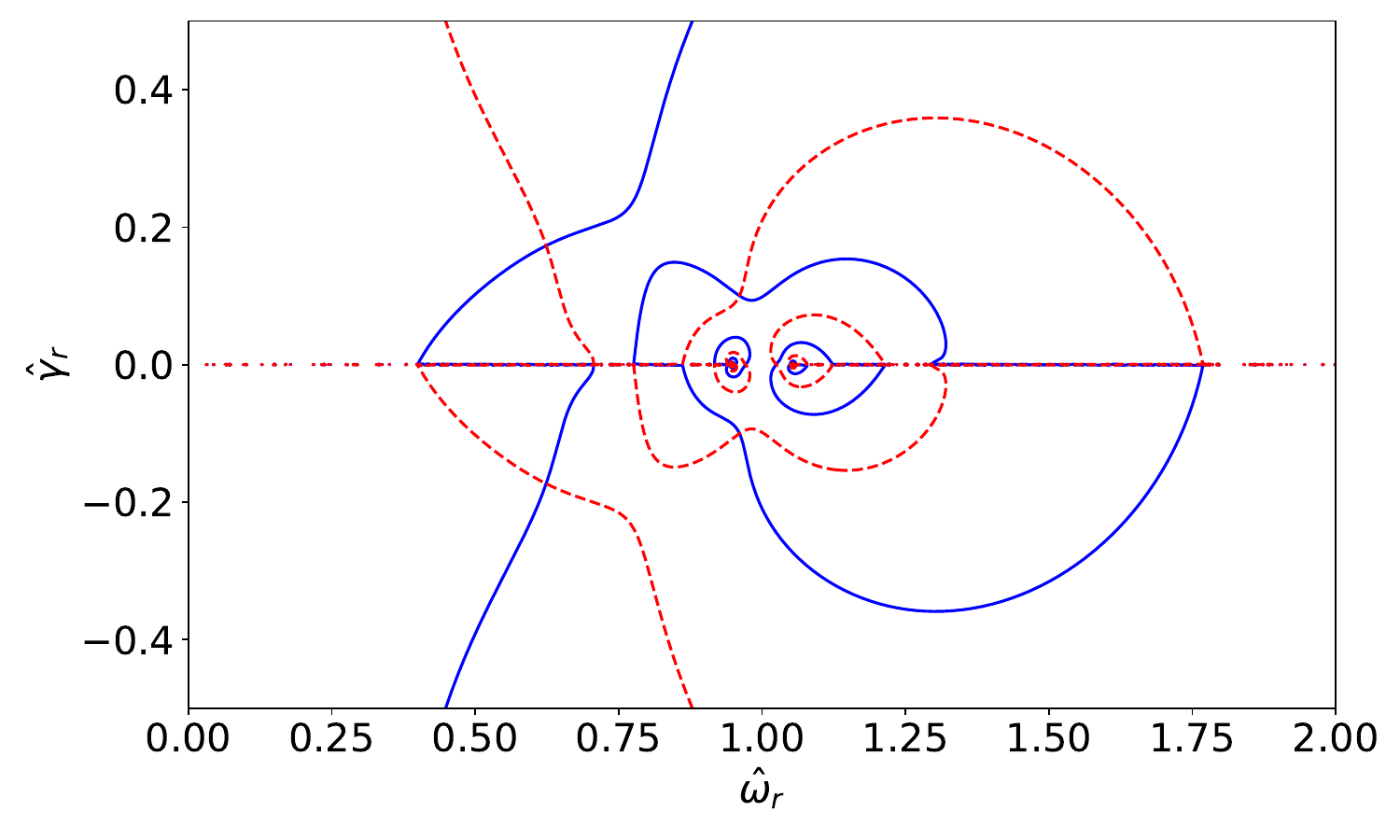}
\caption{Zero contours of ${\rm Re}[\zeta(b)]$ (blue) and ${\rm Im}[\zeta(b)]$  (red) plotted in $\hat{\omega}_r$--$\hat{\gamma}_r$ space  for  an $n=0$, $m=1$, $l=1$ mode in a rigidly rotating plasma characterized by $\hat{b}=3.0$, $\hat{L}=15.0$, and $\hat{\mit\Omega}_\theta=1.0$.  }\label{fig2}
\end{figure}

\begin{figure}
\centering
\includegraphics[width=0.9\textwidth]{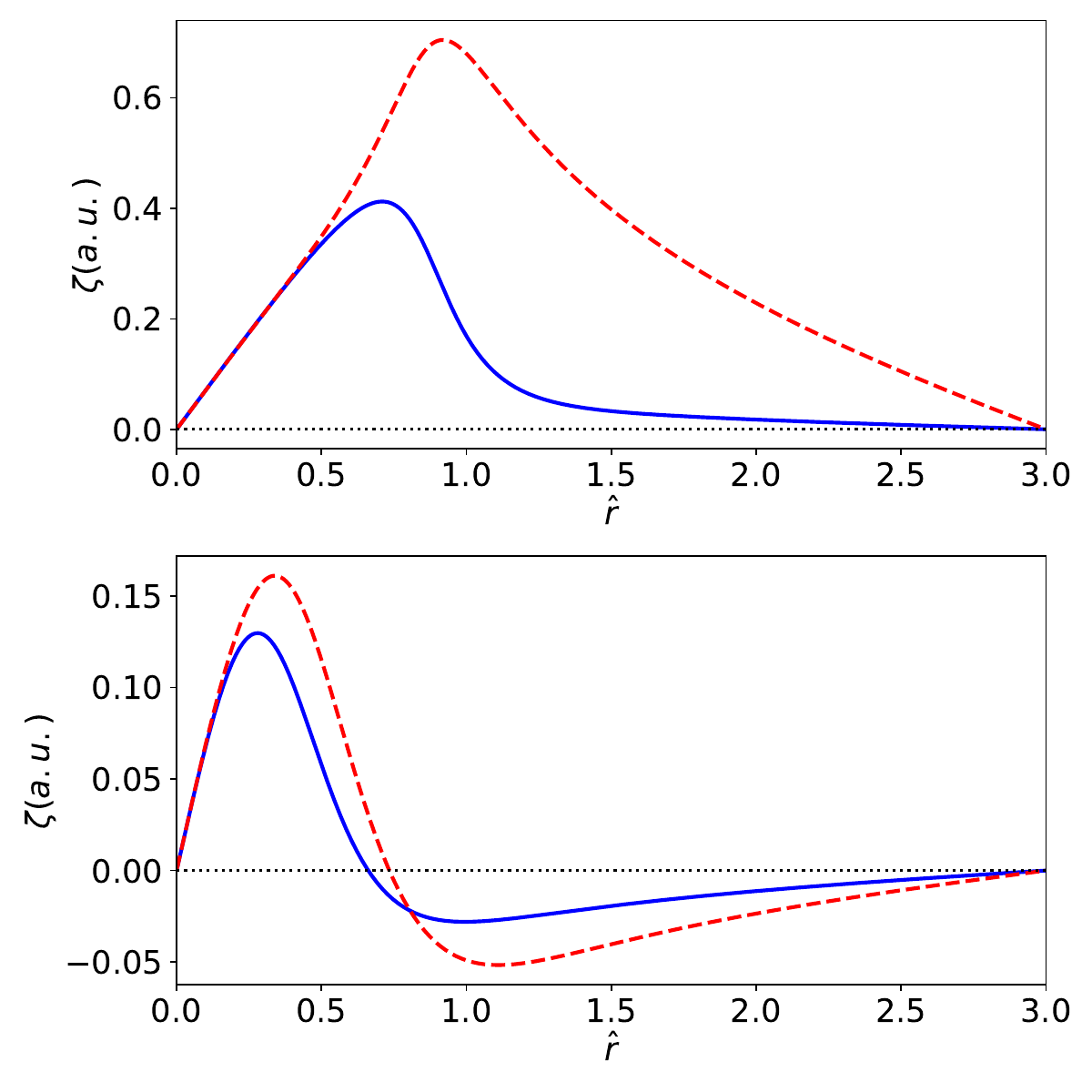}
\caption{Top panel: The eigenfunction of an $n=0$, $m=1$, $l=1$ mode in a rigidly rotating plasma characterized $\hat{b}=3.0$, $\hat{L}=15.0$, and $\hat{\mit\Omega}_\theta=1.0$. The solid blue curve is the real part of the eigenfunction, whereas the dashed red curve is the imaginary part.
Bottom panel: The eigenfunction of an $n=1$, $m=1$, $l=1$ mode in the same plasma equilibrium. }\label{fig3}
\end{figure}

Figure~\ref{fig2} shows the zero contours of ${\rm Re}[\zeta(b)]$ (blue) and ${\rm Im}[\zeta(b)]$ (red), plotted in $\hat{\omega}_r$--$\hat{\gamma}_r$ space 
for an $n=0$, $m=1$, $l=1$
mode in a rigidly rotating plasma characterized by $\hat{b}=3.0$, $\hat{L}=15.0$, and $\hat{\mit\Omega}_\theta=1.0$. It can be seen that there is
a broken  continuum of stable modes lying on the $\hat{\gamma}_r=0$ axis. However, there is also clearly a solitary  unstable mode (where the blue and red curves cross) with normalized
real frequency and growth-rate $\hat{\omega}_r= 0.624$ and $\hat{\gamma}_r= 0.173$, respectively.
Figure~\ref{fig3} displays the eigenfunction of this mode.  The mode in question is clearly an $n=0$ mode, because it only has radial nodes at $\hat{r}=0$ and
$\hat{r}=\hat{b}$.  Note that the
mode does not co-rotate with the plasma, which would correspond to $\hat{\omega}_r=1.0$, but instead rotates more slowly, albeit in the same direction as the plasma. 
It is clear, from Fig.~\ref{fig2}, that there exists a second solitary unstable mode with a higher frequency and a lower growth-rate. 
As shown in Fig.~\ref{fig3}, this mode, which is characterized by
$\hat{\omega}_r= 0.961$ and $\hat{\gamma}_r= 0.100$, is an $n=1$ mode with a single radial node between $\hat{r}=0$ and
$\hat{r}=\hat{b}$.

\begin{figure}
\centering
\includegraphics[width=0.9\textwidth]{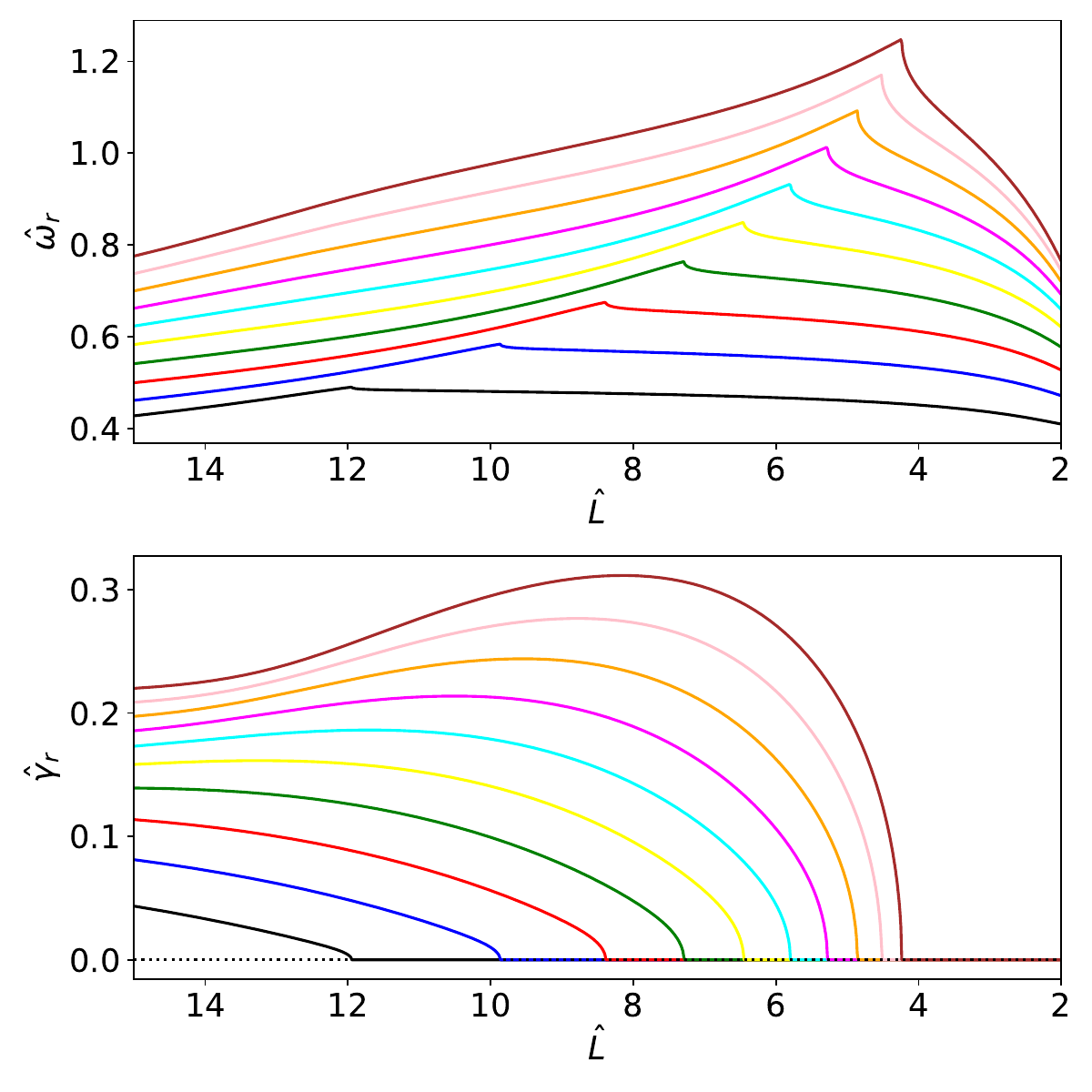}
\caption{The normalized  real frequency  and growth-rate of the $n=0$, $m=1$, $l=1$ mode in a rigidly rotating plasma calculated as  functions of $\hat{L}$ for $\hat{b}=3.0$. The  black, blue, red, green, yellow,  cyan, magenta, orange,  pink, and brown
curves correspond to $\hat{\mit\Omega}_\theta = 0.5$, 0.6, 0.7, 0.8, 0.9, 1.0, 1.1, 1.2, 1.3, and 1.4, respectively.}\label{fig4}
\end{figure}

Figure~\ref{fig4} shows the normalized real frequency and growth-rate of the $n=0$, $m=1$, $l=1$ mode in a rigidly rotating plasma
calculated as  functions of $\hat{L}$ for various different  values of $\hat{\mit\Omega}_\theta$. Note that $\hat{L}\simeq 10$ for the WHAM device [\cite{wham}]. 
It can be seen that the real frequency and growth-rate are both increasing functions of increasing plasma angular velocity. Decreasing the
normalized length, $\hat{L}$, of the mirror has a stabilizing effect on the mode, because of the stabilizing influence of the end-plates. Indeed,
for a sufficiently small value of $\hat{L}$, the $n=0$, $m=1$, $l=1$ mode is completely stabilized. It is clear that WHAM ($\hat{L}=10$) would be stable to
the mode provided $\hat{\mit\Omega}_\theta\lesssim 0.6$. 
Note that the mode becomes progressively harder to stabilize as the plasma rotation increases. 

\subsection{The $n=0$, $m=2$, $l=1$ mode}
\begin{figure}
\centering
\includegraphics[width=0.9\textwidth]{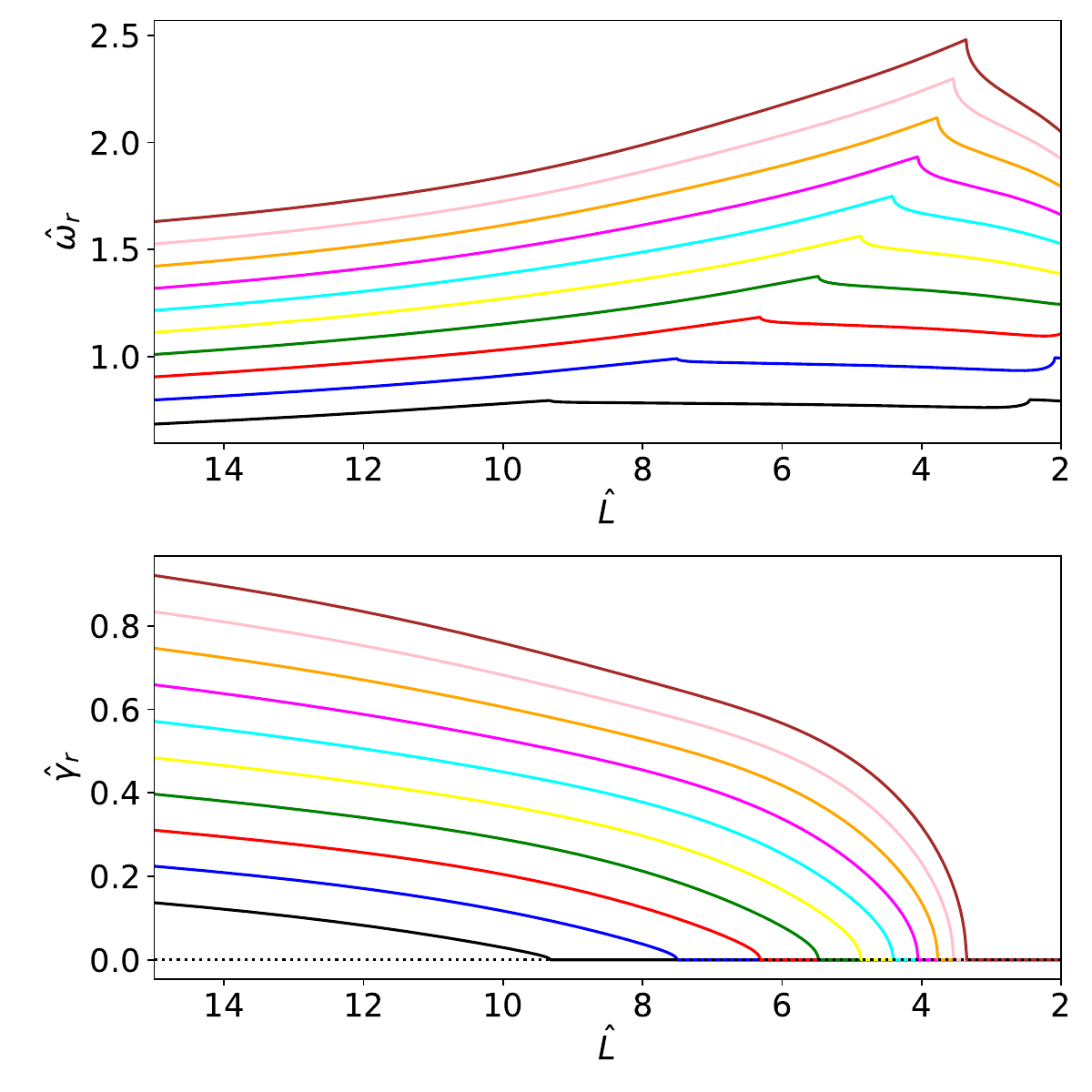}
\caption{The normalized  real frequency  and growth-rate of the $n=0$, $m=2$, $l=1$ mode in a rigidly rotating plasma calculated as  functions of $\hat{L}$ for $\hat{b}=3.0$. The black, blue, red, green, yellow,  cyan, magenta, orange,  pink, and brown
curves correspond to $\hat{\mit\Omega}_\theta = 0.4$, 0.5, 0.6, 0.7, 0.8, 0.9, 1.0, 1.1, 1.2, and 1.3, respectively.}\label{fig5}
\end{figure}

Figure~\ref{fig5} shows the normalized real frequency and growth-rate of the $n=0$, $m=2$, $l=1$ mode in a rigidly rotating plasma
calculated as  functions of $\hat{L}$ for various different values of $\hat{\mit\Omega}_\theta$. It can be seen that this mode generally has a significantly
higher growth-rate than the $n=0$, $m=1$, $l=1$ mode. 
As before, the real frequency and growth-rate are both increasing functions of increasing plasma angular velocity. Decreasing the
normalized length, $\hat{L}$, of the mirror has a stabilizing effect on the mode. However, the critical value of $\hat{L}$ below which stabilization occurs is smaller for this mode than for the corresponding $m=1$ mode. It is clear that WHAM ($\hat{L}=10$) would be stable to
the mode provided $\hat{\mit\Omega}_\theta\lesssim 0.4$. 

\section{Sheared rotation}\label{shear}
Now that we have established a baseline via our exploration of the stability of a rigidly rotating plasma, let us extend our investigation to
take  sheared rotation  into account. 

\subsection{Equilibrium}
Suppose that the plasma does not rotate as a rigid body, so that $\hat{\mit\Omega}_\theta$ is a function of $\hat{r}$. 
Let
\begin{align}\label{eqa}
\hat{N}(\hat{r}) &= \frac{1}{\cosh^2(\hat{r}^{\,2})},\\[0.5ex]
\lambda(\hat{r}) &= \frac{1}{8}\,(\hat{g}+\hat{\mit\Omega}_\theta^{\,2}),\\[0.5ex]
\lambda_0 &= 2\int_0^\infty\lambda(\hat{r})\,\hat{N}(\hat{r})\,\hat{r}\,d\hat{r},\\[0.5ex]
\alpha_0 &= \frac{\lambda_0}{1+\lambda_0},\\[0.5ex]
\beta_0 &= \frac{1}{4}\,(1-\alpha_0)^2.
\end{align}
It follows that
\begin{equation}
\hat{B}_z^{\,2}(\hat{r}) = \alpha_0^{\,2} + 4\,\beta_0\,(1-\hat{N}) + 16\,\beta_0\int_0^{\hat{r}}\,\lambda(\hat{r}')\,\hat{N}(\hat{r}')\,\hat{r}'\,d\hat{r}'.\label{eqb}
\end{equation}
 As before, 
$a$ is a
measure of the width of the current channel, 
$N_0$ the central electron number density, $B_0$  the vacuum magnetic field-strength, and the beta value at the center of the plasma is $\beta_0/\alpha_0^{\,2}$. 

In particular, suppose that
\begin{equation}
\hat{\mit\Omega}_\theta(\hat{r}) = \frac{\hat{\mit\Omega}_{\theta\,0}}{\cosh(\hat{r}/\hat{c})},
\end{equation}
which is a monotonically decreasing, sheared rotation profile that possesses the central angular velocity $\hat{\mit\Omega}_{\theta\,0}$. 
Here, $\hat{c}$ is the normalized shear length. 

\begin{figure}
\centering
\includegraphics[width=0.9\textwidth]{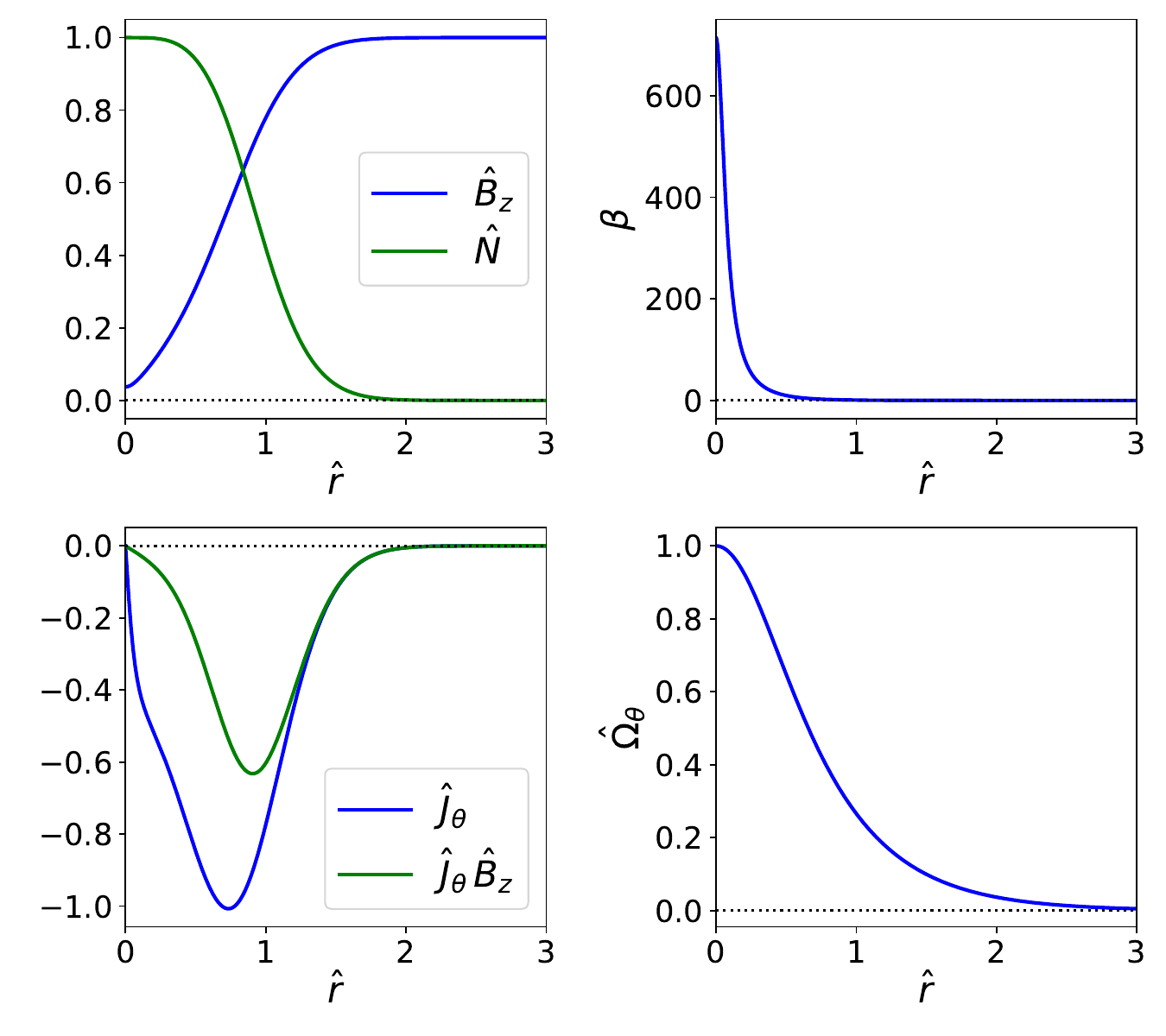}
\caption{A  plasma equilibrium with sheared rotation calculated for $\hat{L}=15.0$, $\hat{\mit\Omega}_{\theta\,0}=1.0$, and $\hat{c}=0.5$. }\label{fig6}
\end{figure}

Figure~\ref{fig6} shows an equilibrium with sheared rotation calculated for $\hat{L}=15.0$, $\hat{\mit\Omega}_{\theta\,0}=1.0$, and $\hat{c}=0.5$. 
The parameters that characterize this equilibrium are $\lambda_0= 0.0389$, $\alpha_0=0.0374$,  $\beta_0= 0.232$, and $\langle \beta\rangle = 0.427$. 

\subsection{The $n=0$, $m=1$, $l=1$ mode}
\begin{figure}
\centering
\includegraphics[width=0.9\textwidth]{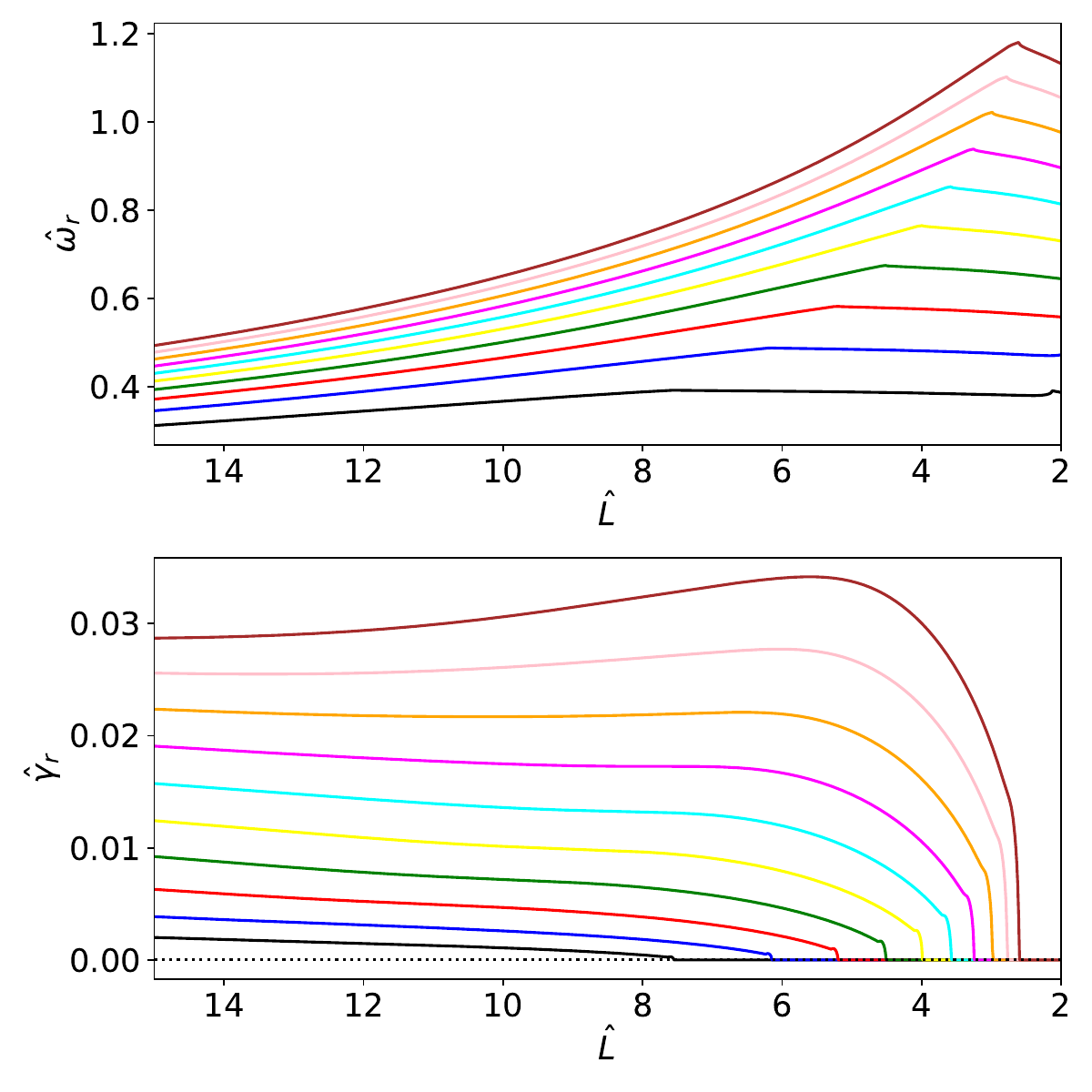}
\caption{The normalized  real frequency  and growth-rate of the $n=0$, $m=1$, $l=1$ mode in a plasma with sheared rotation calculated as  functions of $\hat{L}$ for $\hat{b}=3.0$ and $\hat{c}=0.5$. The  black, blue, red, green, yellow,  cyan, magenta, orange,  pink, and brown
curves correspond to $\hat{\mit\Omega}_{\theta\,0}= 0.4$, 0.5, 0.6, 0.7, 0.8, 0.9, 1.0, 1.1, 1.2, and 1.3, respectively.}\label{fig7}
\end{figure}

Figure~\ref{fig7} shows the normalized real frequency and growth-rate of the $n=0$, $m=1$, $l=1$ mode in a plasma with sheared rotation
calculated as  functions of $\hat{L}$ for various different  values of $\hat{\mit\Omega}_{\theta\,0}$. It can be seen, by comparison with Fig.~\ref{fig4}, that
the real frequency and growth-rate of the mode are less than those in a corresponding rigidly rotating plasma possessing the same central value of
the plasma angular velocity. Indeed, the growth-rate is very much less than in a corresponding rigidly rotating plasma. 
On the other hand, the mode seems harder to completely stabilize in the case with sheared rotation, compared to that with
rigid rotation. To be more exact,  the critical
value of $\hat{L}$ below which the mode is stabilized is smaller in the former case. 

\subsection{The $n=0$, $m=2$, $l=1$ mode}
\begin{figure}
\centering
\includegraphics[width=0.9\textwidth]{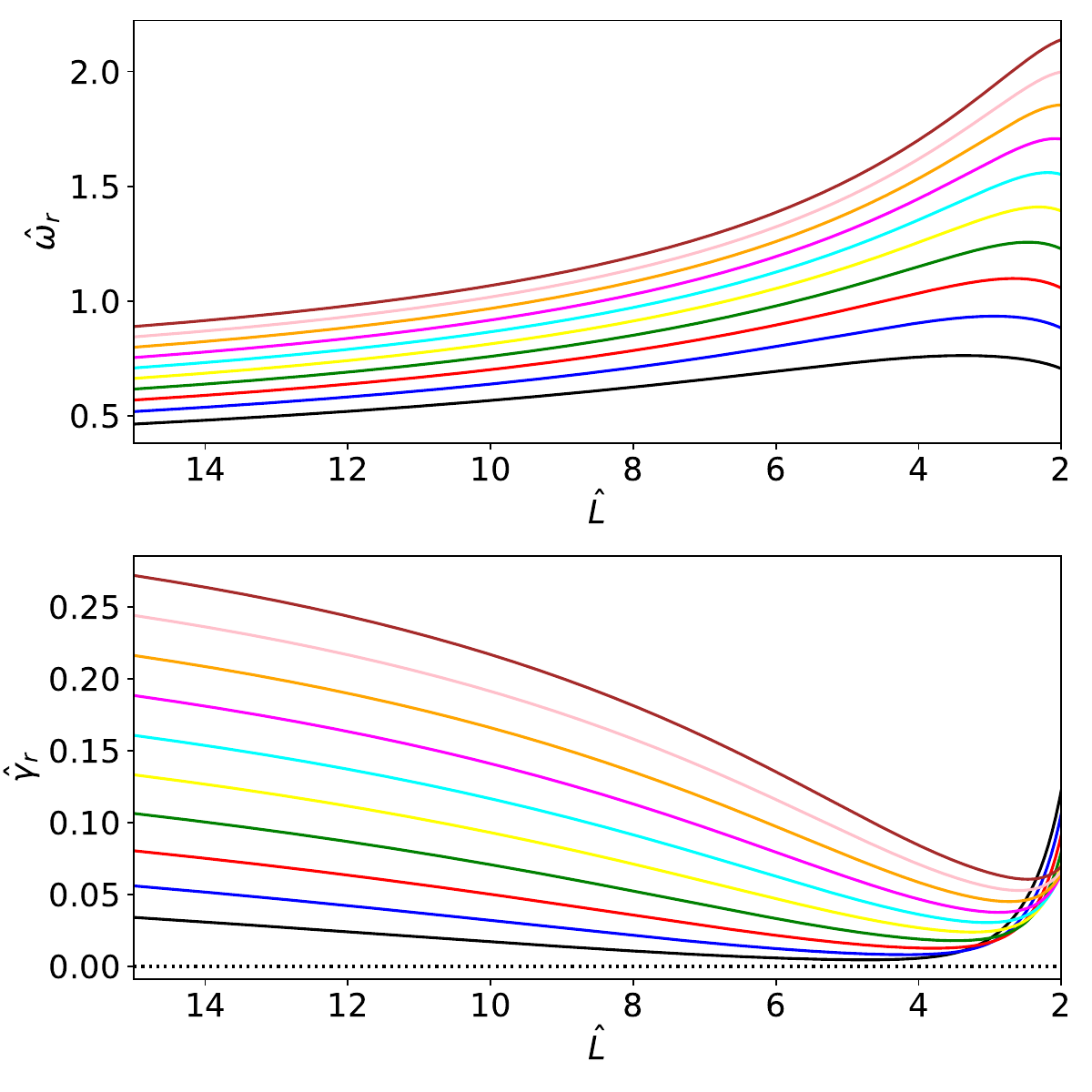}
\caption{The normalized  real frequency  and growth-rate of the $n=0$, $m=2$, $l=1$ mode in a plasma with sheared rotation calculated as  functions of $\hat{L}$ for $\hat{b}=3.0$ and $\hat{c}=0.5$. The  black, blue, red, green, yellow,  cyan, magenta, orange,  pink, and brown
curves correspond to $\hat{\mit\Omega}_{\theta\,0}= 0.4$, 0.5, 0.6, 0.7, 0.8, 0.9, 1.0, 1.1, 1.2, and 1.3, respectively.}\label{fig8}
\end{figure}

Figure~\ref{fig8} shows the normalized real frequency and growth-rate of the $n=0$, $m=2$, $l=1$ mode in a plasma with sheared rotation
calculated as  functions of $\hat{L}$ for various different  values of $\hat{\mit\Omega}_{\theta\,0}$. It can be seen, by comparison with Fig.~\ref{fig5}, that
the real frequency and growth-rate of the mode are less than those in a corresponding rigidly rotating plasma possessing the same central value of
the plasma angular velocity. On the other hand, the mode seems harder to stabilize in the case with sheared rotation, compared to that with
rigid rotation. To be more exact, the mode is always unstable, because the stabilization of the mode as $\hat{L}$ decreases fails at  small
values of $\hat{L}$. Moreover, the growth-rate of the mode is an increasing function of  $\hat{\mit\Omega}_{\theta\,0}$ for intermediate and
large values of $\hat{L}$, but a decreasing function of $\hat{\mit\Omega}_{\theta\,0}$ for small values of $\hat{L}$. 

\section{Vortex flow}\label{vortex}
The main purpose of a magnetic mirror device possessing sonic or supersonic rotation is to take advantage of so-called ``centrifugal confinement'',
by which the centrifugal force acting on charged particles causes them to concentrate on those regions of the  magnetic field-line on
which they are confined that are farthest away from the axis of rotation [\cite{lehnert, bek, hinton, ellis, fitz1}]. Unfortunately, centrifugal confinement does not
work on magnetic-field lines lying  close to the magnetic axis, because these field-lines are almost parallel to the axis. It follows that
there is little benefit to be had by driving strong rotation close to the axis of a mirror device. This observation leads naturally to the idea
of ``vortex flow'', by which the driven rotation is made to peak off the magnetic axis [\cite{bek1}]. Let us investigate the influence of
vortex flow on the macroscopic stability of a rotating theta pinch.

\subsection{Equilibrium}
Our equilibrium is again governed by Eqs.~(\ref{eqa})--(\ref{eqb}). However,  the angular profile now takes the form
\begin{equation}
\hat{\mit\Omega}_\theta(\hat{r}) = 2\,\hat{\mit\Omega}_{\theta\,0}\,\frac{\tanh(\hat{r}/\hat{c})}{\cosh(\hat{r}/\hat{c})}.
\end{equation}
This angular velocity profile is zero on the magnetic axis, and attains it maximum value, $\hat{\mit\Omega}_{\theta\,0}$, at $\hat{r}= 0.8814\,\hat{c}$. 

\begin{figure}
\centering
\includegraphics[width=0.9\textwidth]{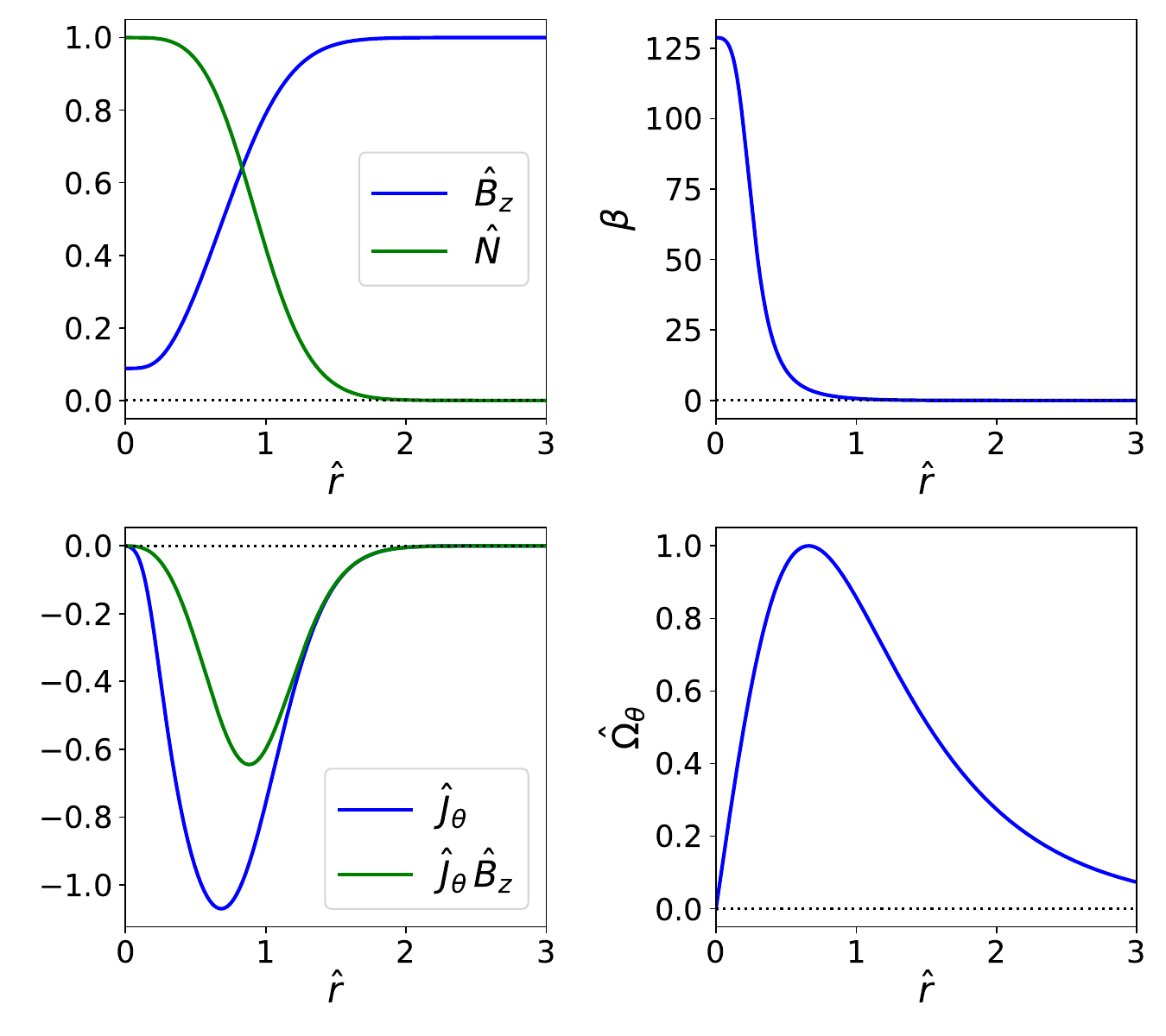}
\caption{A  plasma equilibrium with vortex flow calculated for $\hat{L}=15.0$, $\hat{\mit\Omega}_{\theta\,0}=1.0$, and $\hat{c}=0.75$. }\label{fig9}
\end{figure}

Figure~\ref{fig9} shows an equilibrium with vortex flow calculated for $\hat{L}=15.0$, $\hat{\mit\Omega}_{\theta\,0}=1.0$, and $\hat{c}=0.75$. 
The parameters that characterize this equilibrium are $\lambda_0= 0.0967$, $\alpha_0=0.0881$,  $\beta_0= 0.268$, and $\langle \beta\rangle = 0.330$. 

\subsection{The $n=0$, $m=1$, $l=1$ mode}
\begin{figure}
\centering
\includegraphics[width=0.9\textwidth]{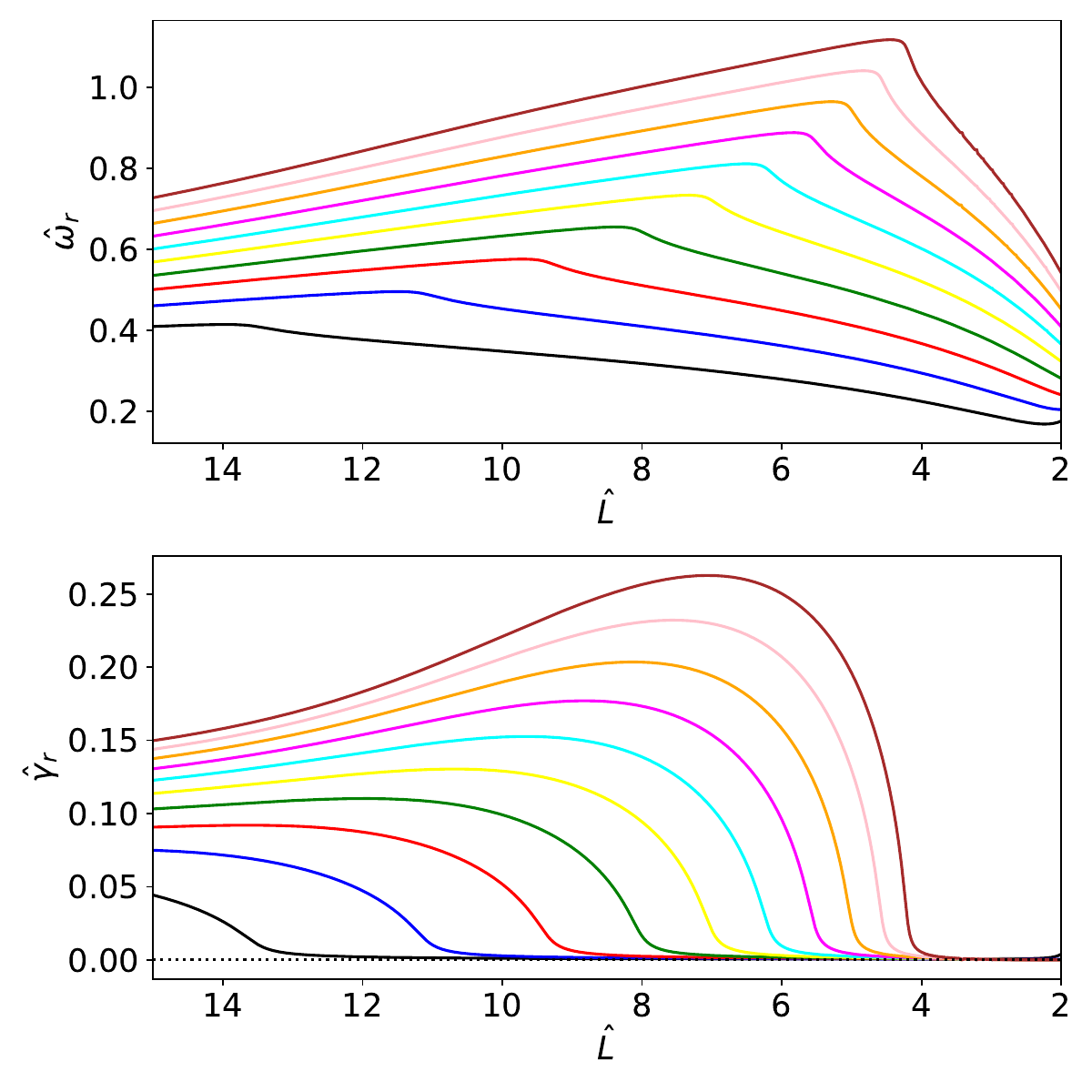}
\caption{The normalized  real frequency  and growth-rate of the $n=0$, $m=1$, $l=1$ mode in a plasma with vortex flow calculated as  functions of $\hat{L}$ for $\hat{b}=3.0$ and $\hat{c}=0.75$. The  black, blue, red, green, yellow,  cyan, magenta, orange,  pink, and brown
curves correspond to $\hat{\mit\Omega}_{\theta\,0}= 0.5$, 0.6, 0.7, 0.8, 0.9, 1.0 1.1, 1.2, 1.3, and 1.4, respectively.}\label{fig10}
\end{figure}

Figure~\ref{fig10} shows the normalized real frequency and growth-rate of the $n=0$, $m=1$, $l=1$ mode in a plasma with vortex flow
calculated as  functions of $\hat{L}$ for various different  values of $\hat{\mit\Omega}_{\theta\,0}$. It can be seen, by comparison with
Fig.~\ref{fig4}, that the  real frequency of the mode is similar to that in the case of rigid rotation, whereas the real
growth-rate a little less. On the other hand, the mode seems a little harder to completely stabilize than in the rigidly rotating case. To be
more exact, as $\hat{L}$ decreases, the growth-rate becomes increasing small, without ever actually becoming zero. 

\subsection{The $n=0$, $m=2$, $l=1$ mode}
\begin{figure}
\centering
\includegraphics[width=0.9\textwidth]{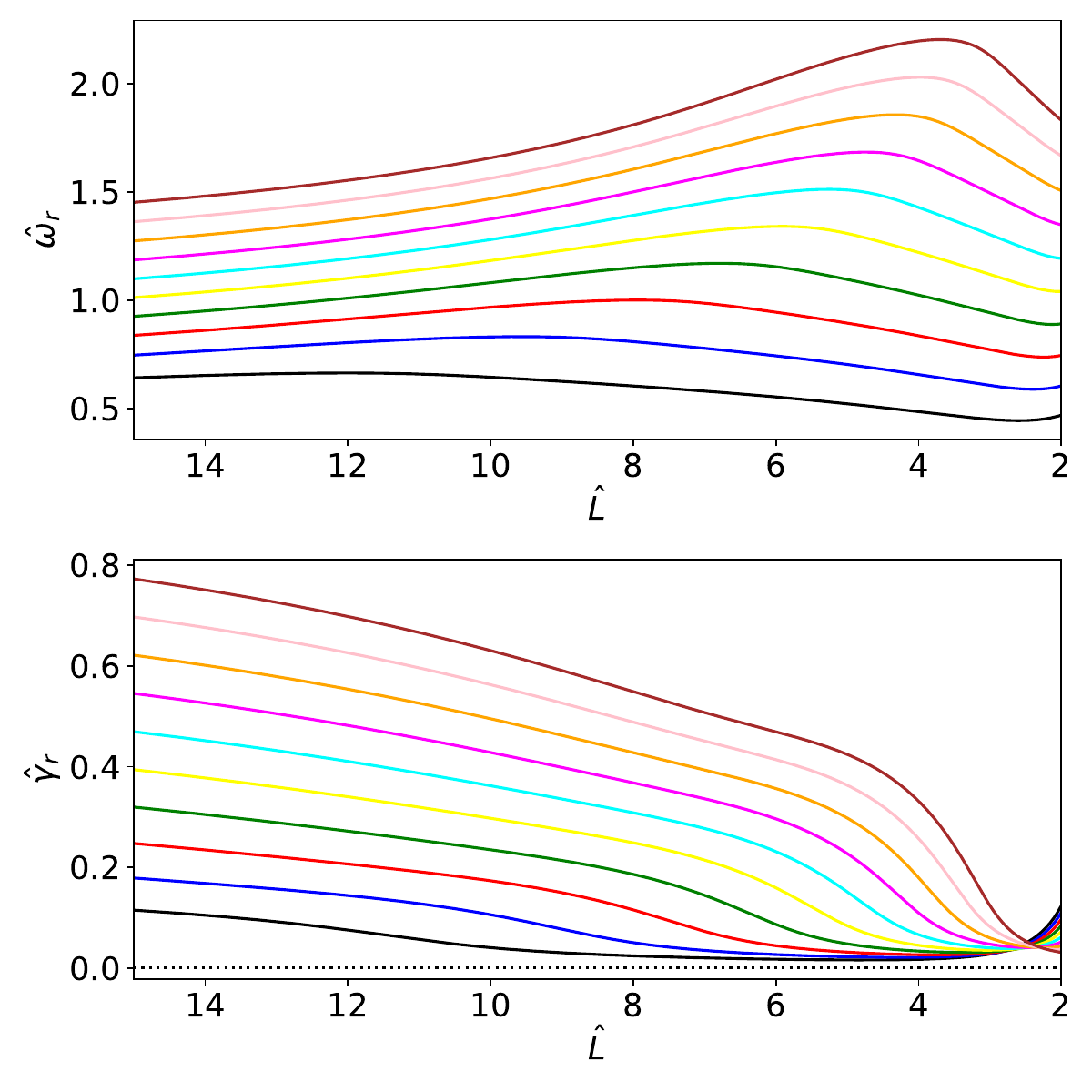}
\caption{The normalized  real frequency  and growth-rate of the $n=0$, $m=2$, $l=1$ mode in a plasma with vortex flow calculated as  functions of $\hat{L}$ for $\hat{b}=3.0$ and $\hat{c}=0.75$. The  black, blue, red, green, yellow,  cyan, magenta, orange,  pink, and brown
curves correspond to $\hat{\mit\Omega}_{\theta\,0}= 0.4$, 0.5, 0.6, 0.7, 0.8, 0.9 1.0, 1.1, 1.2, and 1.3, respectively.}\label{fig11}
\end{figure}

Figure~\ref{fig11} shows the normalized real frequency and growth-rate of the $n=0$, $m=2$, $l=1$ mode in a plasma with vortex flow
calculated as  functions of $\hat{L}$ for various different  values of $\hat{\mit\Omega}_{\theta\,0}$. 
It can be seen, by comparison with Fig.~\ref{fig5}, that the  real frequency and growth-rate of the mode are  similar to that in the case of rigid rotation. 
However, in the case of vortex flow,  the mode is always unstable, because the stabilization of the mode as $\hat{L}$ decreases fails at small
values of $\hat{L}$. Moreover, the growth-rate of the mode is an increasing function of  $\hat{\mit\Omega}_{\theta\,0}$ for intermediate and
large values of $\hat{L}$, but a decreasing function of $\hat{\mit\Omega}_{\theta\,0}$ for small values of $\hat{L}$. 

\section{Diamagnetic effects}\label{dia}
Note that our eigenmode equation, (\ref{eig}),  does not take ion diamagnetic flows into account. The ion diamagnetic frequency is defined [\cite{perl}]
\begin{equation}
{\mit\Omega}_\theta^{\,\ast} = -\frac{m}{e\,N\,B_z\,r}\,\frac{dp_i}{dr},
\end{equation}
where $e$ is the magnitude of the electron charge, and $p_i= N\,T$ the ion pressure. 
Let
${\mit\Omega}_\theta^{\,\ast} = \hat{\mit\Omega}_\theta^{\,\ast}\,v_i/a$. It follows that
\begin{equation}
\hat{\mit\Omega}_\theta^{\,\ast}= - m\,\hat{\rho}_i\,\frac{d\ln\hat{N}}{d\ln\hat{r}},
\end{equation}
where $\hat{\rho}_i=\rho_i/a$, and $\rho_i=v_i/(e\,B_0/m_i)$ is the ion gyro-radius. 
The general criterion for stabilization by diamagnetic effects is [\cite{perl,haines}]
\begin{equation}
|\hat{\gamma}_r| < |\hat{\mit\Omega}_\theta^{\,\ast}|.
\end{equation}
Thus, the previous criterion can only
be satisfied if $|\hat{\gamma}|\lesssim |m|\,\hat{\rho}_i$. Assuming, as seems reasonable, that a well-confined plasma in a
mirror device is characterized by $\hat{\rho}_i\ll 1$ , we conclude that robustly unstable modes 
[i.e., $\hat{\gamma}_r\sim {\cal O}(1)$] are not likely to be stabilized  by
ion diamagnetic effects, unless they possess comparatively large poloidal mode numbers, $|m|$. To be more exact, the only cases that
we have encountered that seem susceptible to diamagnetic stabilization are those in which there is weak off-axis plasma rotation.

\section{Summary and discussion}
In order to gain insight into the macroscopic ideal-MHD stability of an axisymmetric mirror device with sonic levels of rotation, we have approximated
the equilibrium as a rotating theta pinch. The unfavorable curvature of magnetic field-lines in a real mirror is taken into account by introducing
an artificial gravity. 

All of the plasma equilibria investigated in this paper are variants of the rigid rotator equilibrium of \cite{wesson}. As seems reasonable, these equilibria are
characterized by an electron  number density, $N(r)$, that is a monotonically decreasing function of the radial coordinate, $r$. For the sake
of simplicity, we assume that both charge species possess  the same spatially uniform temperature, $T$. 

Employing a single-fluid, ideal-MHD, isothermal fluid model, we have derived a linear, second-order eigenmode equation, (\ref{eig}), for the radial
component of the Lagrangian plasma displacement. This equation
controls the stability of the plasma to small perturbations. In our derivation, we have allowed the plasma angular velocity, ${\mit\Omega}_\theta(r)$,  to
be an arbitrary function of $r$. The first boundary condition imposed on the solution of the eigenmode equation is that it be regular
on the magnetic axis, $r=0$. The second boundary condition  imposed on the solution is
that it be zero at a concentric perfectly-conducting wall that surrounds the plasma at radius $r=b$. The final boundary condition is that the
associated perturbed axial magnetic field be zero at the two perfectly-conducting end-plates that bookend the plasma at $z=0$ and $z=L$. Using our eigenmode equation, we have investigated the
stability of the two most dangerous macroscopic modes: namely, the $m=1$ mode and the $m=2$ mode, where $m$ is the poloidal mode number. 

For the case of a rigidly rotating plasma (see Sect.~\ref{rigid}), we find that both the $m=1$ and the $m=2$ modes are robustly unstable
(i.e., $\hat{\gamma}_r\sim 1$) when the normalized length of the device, $\hat{L}$,  exceeds a critical threshold value. The $m=2$ is more
unstable than the $m=1$ mode. In other words, the former mode possesses a larger growth-rate than the latter, and also possesses a smaller
critical value of $\hat{L}$. As the normalized plasma angular velocity, $\hat{\mit\Omega}_\theta$,  is increased, the
growth-rate of both modes increase, and the critical values of $\hat{L}$ decrease. 

For the case of sheared plasma rotation, in which the plasma angular velocity is largest on the magnetic axis, and decays exponentially
with increasing distance from the axis (see Sect.~\ref{shear}), we find that the real frequencies and growth-rates of the $m=1$ and the
$m=2$ modes are smaller than in a rigidly rotating plasma with the same central angular velocity. Indeed, the real frequency of the $m=1$
is much smaller than in the rigidly rotating case. The growth-rates of both modes increase with increasing central plasma rotation. 
Moreover, the
two modes are harder to completely stabilize than in the rigidly rotating case. In other words, the critical values of $\hat{L}$ are smaller for the $m=1$ mode in the case of sheared rotation. 
Even worse, there is no critical value of $\hat{L}$ below which the $m=2$ mode is stabilized in the case of sheared rotation.   

For the case of vortex flow, in which the plasma angular velocity is zero on the magnetic axis, and peaks off the axis (see Sect.~\ref{vortex}),
the stability of the $m=1$ and $m=2$ modes is broadly similar to that in rigidly rotating plasmas whose angular velocity matches the
peak value. However, the modes are more difficult to completely stabilize than the rigidly rotating case. Indeed, the $m=2$ mode is
always unstable. 

On the whole, our calculations provide little evidence that sheared plasma rotation is an overall stabilizing effect. Indeed, sheared rotation is
capable of reducing the growth-rate of the $m=1$ mode, but seems to preclude stabilization of the
$m=2$ mode. 

There are many possible extensions of the work presented in this paper. We could explore parameter space more fully, looking at a wider range of different
beta values and rotation profiles. We could also investigate the stability of the $m=0$ mode, and $m>2$ modes. Of course, we could relax the isothermal assumption,
allowing for a plasma polytropic index that differs from unity. We could also incorporate  ion diamagnetic effects into the calculation. Finally, we
could incorporate anisotropic temperature into the calculation, because mirror devices generally exhibit higher perpendicular than parallel ion temperatures.

\section*{Funding}
This research was supported by the U.S.\ Department of Energy, Office of Science, Office of Fusion Energy Sciences,  under  contract DE-FG02-04ER54742. 

\section*{Conflicts of interest}
The authors have no conflicts of interest.

\appendix
\section{Derivation of eigenmode equation}\label{secta}
We are assuming an $\exp[\,{\rm i}\,(m\,\theta+\hat{k}\,\hat{z}-\hat{\omega}\,\hat{t})]$ dependence of perturbed quantities. 
Our governing equations, (\ref{e4.1})--(\ref{e4.3d}), yield
\begin{align}\label{eb}
-\hat{k}\,\hat{r}\,{\rm i}\,\delta\hat{B}_z &= (\hat{r}\,\delta\hat{B}_r)' +m\,{\rm i}\,\delta\hat{B}_\theta,
\end{align}
and
\begin{align}
\hat{r}\,\delta\hat{J}_r&=m\,{\rm i}\,\delta\hat{B}_z-\hat{k}\,\hat{r}\,{\rm i}\,\delta\hat{B}_\theta,\label{ec}\\[0.5ex]
{\rm i}\,\delta\hat{J}_\theta &= -\hat{k}\,\delta \hat{B}_ r - {\rm i}\,\delta\hat{B}_z',\label{ed}\\[0.5ex]
\hat{r}\,{\rm i}\,\delta\hat{J}_z&= (\hat{r}\,{\rm i}\,\delta\hat{B}_\theta)' + m\,\delta\hat{B}_r.\label{ef}
\end{align}
and
\begin{align}
\sigma\,\delta\hat{B}_r &= - \hat{k}\,\hat{B}_z\,\delta\hat{V}_r,\label{exx1}\\[0.5ex]
-\sigma\,{\rm i}\,\delta\hat{B}_\theta -\hat{r}\,\hat{\mit\Omega}_\theta'\,\delta\hat{B}_r&= \hat{k}\,\hat{B}_z\,{\rm i}\,\delta\hat{V}_\theta,\label{exx2}\\[0.5ex]
-\sigma\,{\rm i}\,\delta\hat{B}_z&= \hat{k}\,\hat{B}_z \,{\rm i}\,\delta\hat{V}_z -\hat{B}_z'\,\delta\hat{V}_r \nonumber\\[0.5ex]
&\phantom{=}-[(\hat{r}\,\delta\hat{V}_r)'/\hat{r}+m\,{\rm i}\,\delta\hat{V}_\theta/\hat{r}+\hat{k}\,{\rm i}\,\delta\hat{V}_z]\,\hat{B}_z,\label{exxx}
\end{align}
and
\begin{align}\label{en}
\sigma\,{\rm i}\,\delta\hat{N} &= \hat{N}'\,\delta\hat{V}_r + \frac{\hat{N}}{\hat{r}}\,[(\hat{r}\,\delta\hat{V}_r)' + m\,{\rm i}\,\delta\hat{V}_\theta+ \hat{k}\,\hat{r}\,{\rm i}
\,\delta \hat{V}_z]\nonumber\\[0.5ex]
&=\frac{1}{\hat{r}}\left[(\hat{N}\,\hat{r}\,\delta\hat{V}_r)' + m\,{\rm i}\,\hat{N}\,\delta\hat{V}_\theta + \hat{k}\,\hat{r}\,\hat{N}\,{\rm i}\,\delta\hat{V}_z\right],
\end{align}
and
\begin{align}
\hat{B}_z\,{\rm i}\,\delta\hat{J}_\theta+\hat{J}_\theta\,{\rm i}\,\delta\hat{B}_z &= -\beta_0\,\hat{r}\,\hat{\mit\Omega}_\theta^{\,2}\,{\rm i}\,\delta\hat{N}
+2\,\beta_0\,{\rm i}\,\delta\hat{N}'-\beta_0\,\hat{g}\,\hat{r}\,{\rm i}\,\delta\hat{N}\nonumber\\[0.5ex]
&\phantom{=}+\beta_0\,\hat{N}\left(\sigma\,\delta\hat{V}_r -2\,\hat{\mit\Omega}_\theta\,{\rm i}\,\delta\hat{V}_\theta\right),\label{a7}\\[0.5ex]
-\hat{B}_z\,\delta\hat{J}_r &= \frac{2\,\beta_0\,m}{\hat{r}}\,{\rm i}\,\delta\hat{N}
+\beta_0\,\hat{N}\left[-\sigma\,{\rm i}\,\delta\hat{V}_\theta
+(2\,\hat{\mit\Omega}_\theta+ \hat{r}\,\hat{\mit\Omega}_\theta') \,\delta\hat{V}_r\right],\label{a8}\\[0.5ex]
-\hat{J}_\theta\,\delta\hat{B}_r &=2\,\beta_0\,\hat{k}\,{\rm i}\,\delta\hat{N} + \sigma\,\beta_0\,\hat{N}\,{\rm i}\,\delta\hat{V}_z,\label{a9}
\end{align}
respectively. Here, $'\equiv d/d\hat{r}$, and $\sigma(\hat{r})$ is defined in Eq.~(\ref{esig}). 
Note that Eqs.~(\ref{exx1})--(\ref{exxx}) can be combined to give
\begin{equation}
\sigma\, [(\hat{r}\,\delta\hat{B}_r)' +m\,{\rm i}\,\delta\hat{B}_\theta+\hat{k}\,\hat{r}\,{\rm i}\,\delta\hat{B}_z] = 0.
\end{equation}
It follows that Eq.~(\ref{exxx}) is not independent of Eqs.~(\ref{eb}), (\ref{exx1}), and (\ref{exx2}). 

Our closed system of equations is (\ref{eb})--(\ref{ed}), (\ref{exx1}), (\ref{exx2}), (\ref{en})--(\ref{a9}). This constitutes nine equations
involving the nine variables: $\delta\hat{B}_r(\hat{r})$, $\delta\hat{B}_\theta(\hat{r})$, $\delta\hat{B}_z(\hat{r})$, $\delta\hat{J}_r(\hat{r})$, $\delta\hat{J}_\theta(\hat{r})$,
$\delta\hat{N}(\hat{r})$, $\delta\hat{V}_r(\hat{r})$, $\delta\hat{V}_\theta(\hat{r})$, and $\delta\hat{V}_z(\hat{r})$. The remaining variable, $\delta\hat{J}_z(\hat{r})$, is determined independently
from  Eq.~(\ref{ef}). 

Now, Eqs.~(\ref{exx1}), (\ref{exx2}), (\ref{eb}), and (\ref{ec}) give 
\begin{align}
\delta\hat{B}_r& = - \frac{\hat{k}\,\hat{B}_z}{\sigma}\,\delta\hat{V}_r,\\[0.5ex]
{\rm i}\,\delta\hat{B}_\theta &= \frac{\hat{k}\,\hat{B}_z\,\hat{r}\,\hat{\mit\Omega}_\theta'}{\sigma^2}\,\delta\hat{V}_r- \frac{\hat{k}\,\hat{B}_z}{\sigma}\,{\rm i}\,\delta\hat{V}_\theta,\label{e44x}\\[0.5ex]
\hat{r}\,{\rm i}\,\delta\hat{B}_z& = \frac{(\hat{B}_z\,\hat{r}\,\delta\hat{V}_r)' +m\,\hat{B}_z\,{\rm i}\,\delta\hat{V}_\theta}{\sigma},\label{e45x}\\[0.5ex]
\hat{r}^{\,2}\,\delta\hat{J}_r& = \frac{m\,(\hat{B}_z\,\hat{r}\,\delta\hat{V}_r)'}{\sigma} - \frac{(\hat{k}\,\hat{r})^2\,\hat{B}_z
\,\hat{\mit\Omega}_\theta'\,\hat{r}\,\delta\hat{V}_r}{\sigma^2}+ \frac{(m^2+\hat{k}^2\,\hat{r}^{\,2})\,\hat{B}_z\,{\rm i}\,\delta\hat{V}_\theta}{\sigma},\label{jrdef}
\end{align}
respectively. 

It follows from Equation~(\ref{en}) and (\ref{a9}) that 
\begin{equation}\label{e147}
F_s\,\beta_0\,\hat{r}\,{\rm i}\,\delta\hat{N} = -\sigma\,\beta_0\,(\hat{N}\,\hat{r}\,\delta\hat{V}_r)'-\frac{\hat{k}^{\,2}\,\hat{B}_z\,\hat{B}_z'\,\hat{r}\,\delta\hat{V}_r}{\sigma}-\sigma\,\beta_0\,m\,\hat{N}\,{\rm i}\,\delta\hat{V}_\theta,
\end{equation}
where $F_s(\hat{r})$ is defined in Eq.~(\ref{fsdef}). Thus, Eqs.~(\ref{a8}) and (\ref{jrdef}) give 
\begin{align}\label{vtheta}
{\rm i}\,\delta\hat{V}_\theta &= 
-\frac{m\,F_f}{D}\,(\hat{r}\,\delta\hat{V}_r)'
\nonumber\\[0.5ex]
&
\phantom{=}+ \left[\frac{m\,\sigma^2\,\beta\,\hat{r}\,(\hat{g}+\hat{\mit\Omega}_\theta^{\,2})}{D}+ 
\frac{\hat{r}^{\,2}\,\hat{\mit\Omega}_\theta'\,F_a\,F_s}{\sigma\,D} - \frac{2\,\beta\,\sigma\,\hat{r}\,\hat{\mit\Omega}_\theta\,F_s}{D}\right]\hat{r}\,\delta\hat{V}_r, 
\end{align}
where $D(\hat{r})=m^2\,F_f+\hat{r}^{\,2}\,F_a\,F_s$, and $\beta(\hat{r})$, $F_a(\hat{r})$, and $F_f(\hat{r})$ are defined in Eqs.~(\ref{betadef}), (\ref{fadef}), and (\ref{ffdef}), respectively. 

Equations~(\ref{e44x}) and (\ref{vtheta}) yield 
\begin{align}
&\frac{{\rm i}\,\delta \hat{B}_\theta}{\hat{B}_z} = \frac{\hat{k}\,m\,F_f}{D}\,\left(\frac{\hat{r}\,\delta\hat{V}_r}{\sigma}\right)'
-\hat{k}\,\beta\,\hat{r}\left[\frac{m\,\sigma^2\,(\hat{g}+\hat{\mit\Omega}_\theta^{\,2})-2\,\sigma\,\hat{\mit\Omega}_\theta\,F_s}{D}\right]
\frac{\hat{r}\,\delta\hat{V}_r}{\sigma},
\end{align}
whereas Eqs.~(\ref{e45x}) and (\ref{vtheta}) give
\begin{align}\label{bz}
\frac{\hat{r}\,{\rm i}\,\delta\hat{B}_z}{\hat{B}_z}&=
\frac{\hat{r}^{\,2}\,F_a\,F_s}{D}\left(\frac{\hat{r}\,\delta\hat{V}_r}{\sigma}\right)' +\frac{\hat{B}_z'}{\hat{B}_z}\,\frac{\hat{r}\,\delta\hat{V}_r}{\sigma} +m\,\beta\,\hat{r}\left[\frac{m\,\sigma^2\,(\hat{g}+\hat{\mit\Omega}_\theta^{\,2})-2\,\sigma\,\hat{\mit\Omega}_\theta\,F_s}{D}\right]
\frac{\hat{r}\,\delta\hat{V}_r}{\sigma}.
\end{align}

It follows from Eqs.~(\ref{e147}) and (\ref{vtheta}) that 
\begin{align}\label{deltan}
\beta_0\,{\rm i}\,\delta\hat{N}&=-\frac{\sigma^2\,\beta_0\,\hat{N}\,\hat{r}\,F_a}{D}\left(\frac{\hat{r}\,\delta\hat{V}_r}{\sigma}\right)'
-\frac{\sigma\,\beta_0\,F_c\,\hat{N}'}{D}\,\delta\hat{V}_r
\nonumber\\[0.5ex]
&\phantom{=}-
\frac{(m^2+\hat{k}^2\,\hat{r}^{\,2})\,F_a\,\hat{B}_z\,\hat{B}_z'}{\sigma\,D}\,\delta\hat{V}_r+\frac{2\,\beta_0\,\hat{N}\,\beta\,m\,\sigma^2\,\hat{r}\,{\hat\Omega}_\theta}{D}\,\delta\hat{V}_r,
\end{align}
where $F_c(\hat{r})= m^2+\hat{k}^2\,\hat{r}^{\,2} -\beta\,\sigma^2\,\hat{r}^{\,2}$.

Hence, our final equation, (\ref{a7}),  gives 
\begin{align}\label{big}
&\frac{\hat{k}^2\,\hat{B}_z^{\,2}}{\sigma}\,\delta\hat{V}_r-\left[\frac{\hat{B}_z^{\,2}\,\hat{r}\,F_a\,F_s}{D}\left(\frac{\hat{r}\,\delta\hat{V}_r}{\sigma}\right)'\right]'-\left(\frac{\hat{B}_z\,\hat{B}_z'}{\sigma}\,\delta\hat{V}_r\right)'- \left(H\delta\hat{V}_r\right)' =\nonumber\\[0.5ex]
&-\hat{r}\,(\hat{g}+\hat{\Omega}_\theta^{\,2})\left[-\frac{\sigma^2\,\beta_0\,\hat{N}\,\hat{r}\,F_a}{D}\left(\frac{\hat{r}\,\delta\hat{V}_r}{\sigma}\right)'
-\frac{\sigma\,\beta_0\,F_c\,\hat{N}'}{D}\,\delta\hat{V}_r\right.
\nonumber\\[0.5ex]
&\left.-
\frac{(m^2+\hat{k}^2\,\hat{r}^{\,2})\,F_a\,\hat{B}_z\,\hat{B}_z'}{\sigma\,D}\,\delta\hat{V}_r+\frac{2\,\beta_0\,\hat{N}\,\beta\,m\,\sigma^2\,\hat{r}\,{\hat\Omega}_\theta}{D}\,\delta\hat{V}_r\right]\nonumber\\[0.5ex]
&+2
\left[-\frac{\sigma^2\,\beta_0\,\hat{N}\,\hat{r}\,F_a}{D}\left(\frac{\hat{r}\,\delta\hat{V}_r}{\sigma}\right)'
-\frac{\sigma\,\beta_0\,F_c\,\hat{N}'}{D}\,\delta\hat{V}_r\right.
\nonumber\\[0.5ex]
&\left.-
\frac{(m^2+\hat{k}^2\,\hat{r}^{\,2})\,F_a\,\hat{B}_z\,\hat{B}_z'}{\sigma\,D}\,\delta\hat{V}_r+\frac{2\,\beta_0\,\hat{N}\,\beta\,m\,\sigma^2\,\hat{r}\,{\hat\Omega}_\theta}{D}\,\delta\hat{V}_r\right]'\nonumber\\[0.5ex]
&+\beta_0\,\hat{N}\left(\sigma\,\delta\hat{V}_r-2\,\hat{\mit\Omega}_\theta\left\{
-\frac{m\,F_f}{D}\,(\hat{r}\,\delta\hat{V}_r)'\right.\right.\nonumber\\[0.5ex]
&\left.\left.+ \left[\frac{m\,\sigma^2\,\beta\,\hat{r}\,(\hat{g}+\hat{\mit\Omega}_\theta^{\,2})}{D}+ 
\frac{\hat{r}^{\,2}\,\hat{\mit\Omega}_\theta'\,F_a\,F_s}{\sigma\,D} - \frac{2\,\beta\,\sigma\,\hat{r}\,\hat{\mit\Omega}_\theta\,F_s}{D}\right]\hat{r}\,\delta\hat{V}_r\right\}\right),
\end{align}
where
\begin{equation}
H(\hat{r})=  m\,\beta_0\,\hat{N}\,\hat{r}\left[\frac{m\,\sigma\,(\hat{g}+\hat{\mit\Omega}_\theta^{\,2})-2\,\hat{\mit\Omega}_\theta\,F_s}{D}\right],
\end{equation}
and use has been made of Eqs.~(\ref{ed}),  (\ref{vtheta}), (\ref{bz}), and (\ref{deltan}). 
After some analysis, Eq.~(\ref{big}) reduces to the standard form [\cite{goeda}]
\begin{equation}
\frac{d}{d\hat{r}}\!
\left(\frac{N}{D}\,\frac{d\zeta}{d\hat{r}}\right) = \left[A+\frac{B}{D}+\left(\frac{C}{D}\right)'\right]\zeta,
\end{equation}
where
\begin{align}
\zeta(\hat{r}) &= \frac{\hat{r}\,\delta\hat{V}_r}{\sigma},\\[0.5ex]
N(\hat{r}) &= \hat{B}_z^{\,2}\,\hat{r}\,F_a\,F_f,\label{ndef}\\[0.5ex]
A(\hat{r}) &= \frac{\hat{B}_z^{\,2}\,F_a}{\hat{r}}+2\,\beta_0\,\hat{N}\,\hat{\Omega}_\theta\,\hat{\Omega}_\theta'\,\\[0.5ex]
B(\hat{r})&= \beta_0\,\hat{N}'\,(\hat{g}+\hat{\mit\Omega}_\theta^{\,2})\,[(m^2+\hat{k}^2\,\hat{r}^{\,2})\,F_f+\beta\,\sigma^4\,\hat{r}^{\,2}]
\nonumber\\[0.5ex]
&\phantom{=}-\beta_0\,\hat{N}\,\hat{r}\,(\hat{g}+\hat{\mit\Omega}_\theta^{\,2})^2\,(m^2+\hat{k}^2\,\hat{r}^{\,2})\,F_a+ 4\,\beta_0\,\hat{N}\,(\hat{g}+\hat{\Omega}_\theta^{\,2})\,\beta\,m\,\sigma^{3}\,\hat{r}\,\hat{\Omega}_\theta\nonumber\\[0.5ex]
&
\phantom{=}-4\,\beta_0\,\hat{N}\,\beta\,\sigma^2\,\hat{r}\,\hat{\Omega}_\theta^{\,2}\,F_s,\\[0.5ex]
C(\hat{r})&= \beta_0\,\hat{N}\,[\sigma^2\,\hat{r}^{\,2}\,(\hat{g}+\hat{\mit\Omega}_\theta^{\,2})\,F_a+2\,m\,\sigma\,{\mit\Omega}_\theta\,F_f],\\[0.5ex]
D(\hat{r}) &= m^2\,F_f+\hat{r}^{\,2}\,F_a\,F_s.\label{ddef}
\end{align}

\newpage

\newpage

\begin{thebibliography}{}

\bibitem[{\sc Aydemir} (2004)]{aydemir}
{\sc Aydemir}~A.Y.\  2004. {\em Magnetohydrodynamic equilibrium and stability of rotating plasma in a mirror geometry}, Phys.\ Plasmas {\bf 11},
5056.

\bibitem[{\sc Bekhyenev}, et al.\ (1980)]{bek}
{\sc Bekhtenev}~A.A., {\sc Volosov}~V.I., {\sc Pal’chikov}~V.E., {\sc Pekker}~M.S.\ \& {\sc Yudin}~Y.N.\ 1980.
{\em Problems of a thermonuclear reactor with a rotating plasma},
Nucl.\  Fusion {\bf 20}, 579.

\bibitem[{\sc Beklemishev}, et al.\ (2010)]{bek1}
{\sc Beklemishev}~A.D., {\sc Bagryansky}~P.A., {\sc Chaschin}~M.A.\ \& {\sc Soldatkina}~E.I.\ 2010.
 {\em Vortex confinement of plasmas in symmetric mirror traps},
 Fusion Sci.\  \& Technology {\bf 57}, 351.

\bibitem[{\sc Bondeson}, et al.\ (1987)]{bond}
{\sc Bondeson}~A., {\sc Iacono}~R.\ \& {\sc Bhattacharjee}~A.\ 1987. {\em Local magnetohydrodynamic instabilities of cylindrical plasma with sheared equilibrium flows}, Phys.\ Fluids {\bf 30}, 2167.

\bibitem[{\sc Bowers} \& {\sc Haines} (1971)]{haines} 
{\sc Bowers}~E.\ \& {\sc Haines}~M.G.\ 1971.  {\em Application of finite Larmor radius equations for
collisionless plasmas to a $\theta$ pinch}, Phys.\ Fluids {\bf 14}, 165. 

\bibitem[{\sc Chandrasekhar} (1967)]{chand}
{\sc Chandrasekhar}~A.\ 1967.\ {\em An introduction to the study of stellar structure}. (Dover, New York NY, 1967).

\bibitem[{\sc Ellis}, et al.\ (2001)]{ellis}
{\sc Ellis}~R.F., {\sc Hassam}~A.B.,  {\sc Messer}~S.\ \&   {\sc Osborn}~B.\ 2001. 
{\em An experiment to test centrifugal confinement for fusion},
Phys.\  Plasmas {\bf 8}, 2057.

\bibitem[{\sc Endrizzi}, et al.\ (2023)]{wham}
{\sc Endrizzi}~D., {\sc Anderson}~J.K., {\sc Brown}~M., {\sc Egedal}~J., {\sc Geiger}~B., {\sc Harvey}~R.W., {\sc Ialovega}~M., {\sc Kirch}~J., 
{\sc Peterson}~E., {\sc Petrov}~Y.,  {\sc Pizzo}~J., 
{\sc Qian}~T., {\sc Sanwalka}~K.,  {\sc Schmitz}~O.,  {\sc Wallace}~J., {\sc Yakovlev}~D., {\sc Yu}~M.\ \& {\sc Forest}~C.B.\ 2023.  {\em Physics basis for the Wisconsin HTS Axisymmetric Mirror (WHAM)}, J.\ Plasma Phys.\ {\bf 89}, 975890501.

\bibitem[{\sc Ferraro} (1937)]{ferraro}
{\sc Ferraro}~V.C.A.\ 1937.\  {\em The non-uniform rotation of the Sun and its magnetic ﬁeld},  Mon.\ Not.\ R.\ Astron.\ Soc.\ {\bf 97}, 458.

\bibitem[{\sc Fitzpatrick} (2023)]{fitz}
{\sc Fitzpatrick}~R.\ 2023.  {\em Plasma physics: An introduction}, 2nd Ed.\ (Taylor \& Francis Group, CRC Press, Baca Raton FL, 2023).

\bibitem[{\sc Fitzpatrick} (2023a)]{fitz1a}
{\sc Fitzpatrick}~R.  2023a. {\em Tearing mode dynamics in tokamak plasmas}. (IOP Publishing, Bristol UK, 2023).

\bibitem[{\sc Fitzpatrick} (2026)]{fitz1}
{\sc Fitzpatrick}~R.  2026. {\em Equilibrium of a rapidly rotating axisymmetric magnetic mirror machine}, arXiv:2607.18203, to appear in J.\ Plasma
Phys.

\bibitem[{\sc Freidberg} (2014)]{freidberg}
{\sc Freidberg}~J.P.\  2014.  {\em Ideal MHD}. (Cambridge University Press, Cambridge UK, 2014). 

\bibitem[{\sc Freidberg} \& {\sc Pearlstein} (1978)]{perl}
{\sc Friedberg}~J.P.\ \& {\sc Pearlstein}~L.D.\  1987. {\em Rotational instabilities in a theta pinch}, Phys.\ Fluids {\bf 21}, 1207.

\bibitem[{\sc Freidberg} \& {\sc Wesson} (1970)]{wesson} 
{\sc Freidberg}~J.P.\ \& {\sc Wesson}~J.A.\  1970. {\em Instability of the $m=1$ mode of a rotating $\theta$ pinch}, Phys.\ Fluids {\bf 13}, 1117.

\bibitem[{\sc Frieman} \& {\sc Rotenberg} (1960)]{frie} 
{\sc Frieman}~E. \& {\sc Rotenberg}~M.\ 1960,
{\em On hydromagnetic stability of stationary equilibria}, 
Rev.\ Mod.\ Phys.\ {\bf 32}, 898.

\bibitem[{\sc Goedbloed} (2018)]{goed}
{\sc Goedbloed}~J.P.\ 2018. {\em The Spectral Web of stationary plasma equilibria. I. General theory}, Phys.\ Plasmas {\bf 25}, 032109.

\bibitem[{\sc Goedbloed} (2018a)]{goeda}
{\sc Goedbloed}~J.P.\ 2018a. {\em The Spectral Web of stationary plasma equilibria. II. Internal modes}, Phys.\ Plasmas {\bf 25}, 032110.


\bibitem[{\sc Hinton} \& {\sc Wong} (1985)]{hinton}
{\sc Hinton}~F.L.\ \& {\sc Wong}~S.K.\ 1985. {\em Neoclassical ion transport in rotating axisymmetric plasmas}, Phys.\ Fluids {\bf 28}, 3028.

\bibitem[{\sc Hojo} (2010)]{hojo}
{\sc Hojo}~H.\ 2010.  {\em Flute-mode stability of quadrupole-anchored tandem mirror plasmas},  Plasma Fusion Research {\bf 5}, 008. 

\bibitem[{\sc Ioffe}, et al.\  (1958)]{ioffe}
 {\sc Ioffe}~M.S., {\sc Telkovskii}~V.G.,  {\sc Yushmanov}~V.K.\ \&  {\sc Sobolev}~E.I.\ 1958. 
{\em Hydromagnetic stability of a high-temperature plasma in a magnetic trap}, 
Proceedings of the Second United Nations International Conference on the Peaceful Uses of Atomic Energy, Geneva, Vol. 32, pp. 799–806.

\bibitem[{\sc Lehnert} (1971)]{lehnert}
{\sc Lehnert}~B.\ 1971. {\em Rotating plasmas},  Nucl.\ Fusion {\bf 11}, 485.

\bibitem[{\sc Morse} \& {\sc Freidberg} (1970)]{morse}
{\sc Morse}~R.L.\ \& {\sc Freidberg}~J.P.\ 1970.
{\em Rigid drift model of high-temperature plasma containment},
Phys.\ Fluids {\bf 13}, 531. 

\bibitem[{\sc Post} (1987)]{post}
{\sc Post}~R.E.\ 1987. {\em The magnetic mirror approach to fusion}, Nucl.\ Fusion {\bf 27}, 1589. 

\bibitem[{\sc Rosenbluth} \& {\sc Longmuir} (1957)]{rosenbluth}
{\sc Rosenbluth}~M.N.\ \&  {\sc Longmire}~C.L.\ 1957. 
{\em Stability of plasmas confined by magnetic fields},
Annals of Physics (N.Y.) {\bf 1}, 120.

\bibitem[{\sc Rosenbluth} \& {\sc Rostoker} (1959)]{rosen}
{\sc Rosenbluth}~M.N.\ \& {\sc Rostoker}~N.\ 1959.  {\em Theoretical structure of plasma equations},  Phys.\ Fluids {\bf 2}, 23 (1959). 

\bibitem[\sc{Ryutov} (1990)]{ryutov}
{\sc Ryutov}~D.D.\ 1980.  {\em Mirror devices}, Plasma Devices and Operations {\bf 1}, 79.  

\end{thebibliography}
\end{document}